\newcommand{\buildtarget}{arxiv}

\documentclass[letterpaper, 10 pt, conference]{ieeeconf}  

\IEEEoverridecommandlockouts                              

\usepackage{cite}
\usepackage{graphicx} 
\usepackage[utf8]{inputenc}
\usepackage{amsmath}  
\usepackage{amssymb}  
\usepackage{array}  
\usepackage[dvipsnames]{xcolor}  
\usepackage{algorithm}
\usepackage{algorithmicx}
\usepackage{algpseudocode}
\usepackage[colorlinks,bookmarksopen,bookmarksnumbered,citecolor=red,urlcolor=blue]{hyperref}
\usepackage{tcolorbox}
\tcbuselibrary{breakable,skins}
\usepackage{bm}
\usepackage{subcaption}
\usepackage{makecell}
\usepackage[table]{xcolor}

\newtheorem{thm}{Theorem}
\newtheorem{defn}{Definition}

\usepackage{etoolbox}
\makeatletter
\patchcmd{\@thm}{\@kernel@refstepcounter}{\refstepcounter}{}{}
\makeatother

\newcommand{\appendixref}[1]{\hyperref[#1]{Appendix~\ref*{#1}}}

\usepackage{tikz}
\usetikzlibrary{arrows.meta}
\usepackage{pgfplots}
\pgfplotsset{compat=newest}
\usepgfplotslibrary{groupplots}
\usepackage{booktabs}
\usepackage{array}
\newcolumntype{C}[1]{>{\centering\arraybackslash}m{#1}} 
\newcommand{\edits}[1]{\textcolor{black}{#1}}

\DeclareMathOperator*{\argmax}{arg\,max}

\usepackage[nolist]{acronym}
\DeclareRobustCommand{\acitalic}[1]{%
  \begingroup
    \renewcommand*{\acffont}[1]{\textit{##1}}%
    \renewcommand*{\acfsfont}[1]{\textnormal{##1}}%
    \ac{#1}%
  \endgroup
}
\acrodef{MPC}{model predictive control}
\acrodef{CBF}{control barrier function}
\acrodef{CLF}{control Lyapunov function}
\acrodef{HJ}{Hamilton-Jacobi}
\acrodef{GP}{Gaussian processe} 
\acrodef{NN}{neural networks}
\acrodef{BLR}{Bayesian linear regression}

\acrodef{ISS}{input-to-state stability}
\acrodef{dISS}[$\delta$-ISS]{incremental input-to-state stability}

\acrodef{PSF}{predictive safety filter}
\acrodef{CMDP}{constrained Markov decision processe}

\acrodef{LQR}{linear-quadratic regulator}
\acrodef{LTI}{linear time-invariant}
\acrodef{PE}{persistently exciting}
\acrodef{QP}{quadratic program}
\acrodef{DeePC}{data-enabled predictive control} 

\acrodef{RL}{reinforcement learning}
\acrodef{DRL}{deep reinforcement learning}
\acrodef{BO}{Bayesian optimization}
\acrodef{IL}{imitation learning}

\acrodef{ETL}{event-triggered learning}
\acrodef{ODE}{ordinary differential equation}
\acrodef{iid}[i.i.d.]{independent and identically distributed}

\acrodef{DQN}{deep Q-network}
\acrodef{A3C}{asynchronous advantage actor-critic}
\acrodef{TRPO}{trust-region policy optimization}
\acrodef{PPO}{proximal policy optimization}
\acrodef{TD3}{twin-delayed deep deterministic policy gradient}
\acrodef{TDMPC}[TD-MPC]{temporal-difference MPC}

\acrodef{CDC}{Conference on Decision and Control}
\acrodef{ACC}{American Control Conference}
\acrodef{ECC}{European Control Conference}

\usepackage{fontawesome5}
\usepackage{multicol}
\tcbuselibrary{skins,breakable}

\definecolor{tutorialblue}{RGB}{35,105,176}

\newcommand{\tutorialbodyfont}{\small}

\newcommand{\tutorialtitlelabel}{Tutorial Example and Implementation}
\newcommand{\tutorialcodelabel}{Code}

\newcommand{\codelink}[1]{%
  \href{#1}{\textcolor{tutorialblue}{\faIcon{external-link-alt}}}%
}

\usepackage{etoolbox}
\newif\iftutorialcodelink
\providecommand{\buildtarget}{arxiv}
\ifdefstring{\buildtarget}{arxiv}{\tutorialcodelinktrue}{%
  \ifdefstring{\buildtarget}{cdc_camera_ready}{\tutorialcodelinkfalse}{%
    \PackageError{exampleboxsetup}%
      {Unknown build target '\buildtarget'}%
      {Set \string\buildtarget\space to 'arxiv' or 'cdc_camera_ready' at the
       top of root.tex.}%
  }%
}

\newcommand{\tutoriallinksep}{\quad}
\makeatletter
\newcommand{\titlecodelink}{%
  \@ifstar
    {\def\tutorial@linkglue{\tutoriallinksep}\tutorial@titlelink}%
    {\def\tutorial@linkglue{\hspace{0pt plus 1filll}}\tutorial@titlelink}%
}
\newcommand{\tutorial@titlelink}[2][\tutorialcodelabel]{%
  \iftutorialcodelink
    \tutorial@linkglue
    \href{#2}{\textcolor{white}{\mdseries\small #1~\faIcon{external-link-alt}}}%
  \fi
}
\makeatother

\newcommand{\tutorialheading}[1]{\noindent\textbf{#1.}\enspace\ignorespaces}

\makeatletter
\newcommand*{\tcb@cnt@tutorialexampleboxautorefname}{Example}

\makeatother

\tcbset{
  tutorialstyle/.style={
    enhanced jigsaw,
    colback=tutorialblue!4!white,
    colframe=tutorialblue,
    colbacktitle=tutorialblue,
    coltitle=white,
    fonttitle=\bfseries,
    fontupper=\tutorialbodyfont,
    boxrule=0.5pt,
    arc=1mm,
    left=5pt,
    right=5pt,
    top=4pt,
    bottom=4pt,
    toptitle=1pt,
    bottomtitle=1pt,
    before skip=0pt,
    after skip=0pt,
    before upper={\setlength{\parindent}{1em}\setlength{\parskip}{0pt}},
  }
}

\newtcolorbox[auto counter]{tutorialexamplebox}[3][]{%
  tutorialstyle,
  breakable,
  lines before break=4,
  width=\columnwidth,
  before skip=6pt plus 2pt minus 1pt,
  after skip=6pt plus 2pt minus 1pt,
  title={\tutorialtitlelabel~\thetcbcounter\titlecodelink{#3}},
  title after break={\tutorialtitlelabel~\thetcbcounter\ (continued)},
  before upper app={\tutorialheading{#2}},
  #1
}

\newtcolorbox[use counter from=tutorialexamplebox]{tutorialexamplewidebox}[2][]{%
  tutorialstyle,
  width=\textwidth,
  title={\tutorialtitlelabel~\thetcbcounter\titlecodelink{#2}},
  #1
}

\makeatletter
\newif\iftutorial@cols
\newcommand{\tutorial@begincols}{\begin{multicols}{2}\global\tutorial@colstrue}
\newcommand{\tutorial@endcols}{\end{multicols}\global\tutorial@colsfalse}

\newenvironment{tutorialexample*}[3][]{%
  \begingroup
    \tcbset{floatplacement=tp,#1}%
    \edef\tutorial@tmp{\endgroup\noexpand\begin{figure*}[\kvtcb@floatplacement]}%
  \tutorial@tmp
  \setlength{\multicolsep}{3pt plus 1pt minus 1pt}%
  \setlength{\columnseprule}{0pt}%
  \raggedcolumns
  \begin{tutorialexamplewidebox}[#1]{#3}%
  \tutorial@begincols
  \tutorialheading{#2}%
}{%
  \iftutorial@cols\tutorial@endcols\fi
  \end{tutorialexamplewidebox}%
  \end{figure*}%
}

\newcommand{\tutorialwidefigure}{%
  \@ifstar{\tutorial@widefigure{}}{\tutorial@widefigure{\tutorial@begincols}}%
}
\newcommand{\tutorial@widefigure}[2]{%
  \iftutorial@cols\tutorial@endcols\fi
  \par\addvspace{\medskipamount}%
  \noindent\begin{minipage}{\linewidth}\centering #2\end{minipage}\par
  \addvspace{\medskipamount}%
  #1%
}
\makeatother

\makeatletter
\newcommand{\tutorialcolumnbreak}{\iftutorial@cols\columnbreak\fi}
\makeatother

\newenvironment{tutorialfigure}
  {\par\medskip\noindent
   \begin{minipage}{\linewidth}\centering}
  {\end{minipage}\par\medskip}

\usepackage{multirow}

\title{\LARGE \bf
A Decade of Learning-Based Control: \\From  Theoretic Foundations to Practical Deployment
}

\author{Angela P. Schoellig$^1$,  Matthias A. Müller$^2$, Sebastian Trimpe$^3$, Melanie N. Zeilinger$^4$,\\Mahathi Anand$^1$, Andrea Carron$^4$, Oliver Hausdörfer$^1$,\\Victor G. Lopez$^2$, Friedrich Solowjow$^3$, Haocheng Zhao$^1$, and Siqi Zhou$^5$
\thanks{$^1$Chair of Safety, Performance, and Reliability for Learning Systems, Technical University of Munich}
\thanks{$^2$Institute of Automatic Control, Leibniz University Hannover}
\thanks{$^3$Institute for Data Science in Mechanical Engineering, RWTH Aachen University}
\thanks{$^4$Institute for Dynamical Systems and Control, ETH Zurich}
\thanks{$^5$School of Computing Science, Simon Fraser University}
\thanks{Angela P. Schoellig,  Matthias A. Müller,  Sebastian Trimpe, and Melanie N. Zeilinger are the organizers of the CDC Learning-Based Control Invited Sessions and lead authors of this paper. Siqi Zhou led the coordination and development of the manuscript. Mahathi Anand, Andrea Carron, Oliver Hausdörfer, Victor G.~Lopez,  Friedrich Solowjow, and Haocheng Zhao (in alphabetical order) contributed to the paper.}
}

\begin{document}

\maketitle
\thispagestyle{empty}
\pagestyle{empty}

\begin{abstract}

Motivated by ten years of Learning-Based Control Invited Sessions at the IEEE Conference on Decision and Control (CDC), this tutorial paper reviews key advances at the intersection of machine learning and control. We focus on methods that combine the flexibility of machine learning techniques with the structure and theoretical foundations of control to design autonomous systems that are capable, reliable, and safe. We first introduce core notions for establishing stability and safety guarantees, then review approaches that learn dynamics models to improve control under uncertainty, and finally discuss data-driven methods that directly synthesize or learn policies, including reinforcement learning, imitation learning, and Bayesian optimization. We further discuss representative applications and open challenges. The goal is to provide an accessible entry point to the field and to support continued progress toward reliable learning-based control in real-world systems. \edits{To support learning and practical exploration, we share implementations and illustrative examples along with our discussion, together with additional resources on our website: \href{https://learning-based-control.github.io/tutorial/}{https://learning-based-control.github.io/tutorial/}.}
\end{abstract}

\section{Introduction}
Over the past decade, the integration of machine learning and control theory has evolved from a growing research interest into a practical approach for many real-world systems. Machine learning provides powerful tools for extracting patterns from data, while control theory offers the mathematical foundation needed to use such data-driven methods in closed-loop systems. Principled combinations of these perspectives can extend autonomous systems to more challenging operating domains, broader system classes, and more complex tasks, while preserving the closed-loop properties required for reliable deployment.

\textit{Why machine learning?} With the widespread availability of data and accelerated computing, machine learning methods have achieved remarkable success across many domains, especially in computer vision and natural language processing. From early milestones in image recognition~\cite{deng2009imagenet, krizhevsky2012imagenet} to recent large-scale language models~\cite{vaswani2017attention,achiam2023gpt,touvron2023llama}, these methods have demonstrated a strong capability to capture complex input-output relationships directly from unstructured data. A similar shift is taking place in control through deep \ac{RL}. Enabled by high-quality synthetic data and massively parallelized compute power~\cite{mittal2023orbit,todorov2012mujoco,schuck2026crazyflow}, deep \ac{RL} has expanded our ability to address challenging control problems involving high-dimensional, non-smooth, and complex dynamical systems~\cite{rudin2022learning, kaufmann2023champion}. More broadly, leveraging data, whether synthetic or real, offers new opportunities to expand what autonomous systems can achieve.

\begin{figure}[t]
    \centering
\includegraphics[width=\columnwidth]{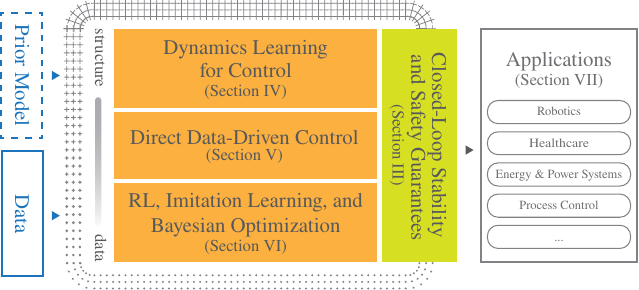}
    \caption{Scope of the tutorial paper. We review learning-based control methods that combine machine learning with control-theoretic structure to improve system capabilities while preserving desired closed-loop properties. The paper first introduces foundations for stability and safety guarantees (green), then covers methodological paradigms ranging from structured model-based learning to increasingly data-driven policy learning (orange), and finally discusses representative application domains (grey).
}
    \label{fig:overview}
\end{figure}
\textit{Why control?} The flexibility that makes machine learning powerful also introduces challenges when learning-based methods are deployed in physical  systems. Purely data-driven models can be highly expressive, but often struggle to satisfy strict safety and reliability requirements, and may require large amounts of data that are costly or infeasible to collect on real systems. Control provides structural priors and theoretical tools for bridging this gap. By embedding control principles into the learning design, one can guide the learning process toward methods that are more sample efficient, high-performing, and safe for practical applications~\cite{safelearn,hewing2020learning,bansal2017hamilton,ames2019control}. The use of data to model the unknown has long been central to control, for instance, in system identification and adaptive control. However, the scale, expressiveness, and opacity of modern learning methods create new challenges in sample efficiency, verification, and reliable deployment. Addressing these challenges requires a tight and principled coupling of machine learning and control-theoretic approaches.

\begin{figure}[t]
    \centering
    \includegraphics[width=\columnwidth]{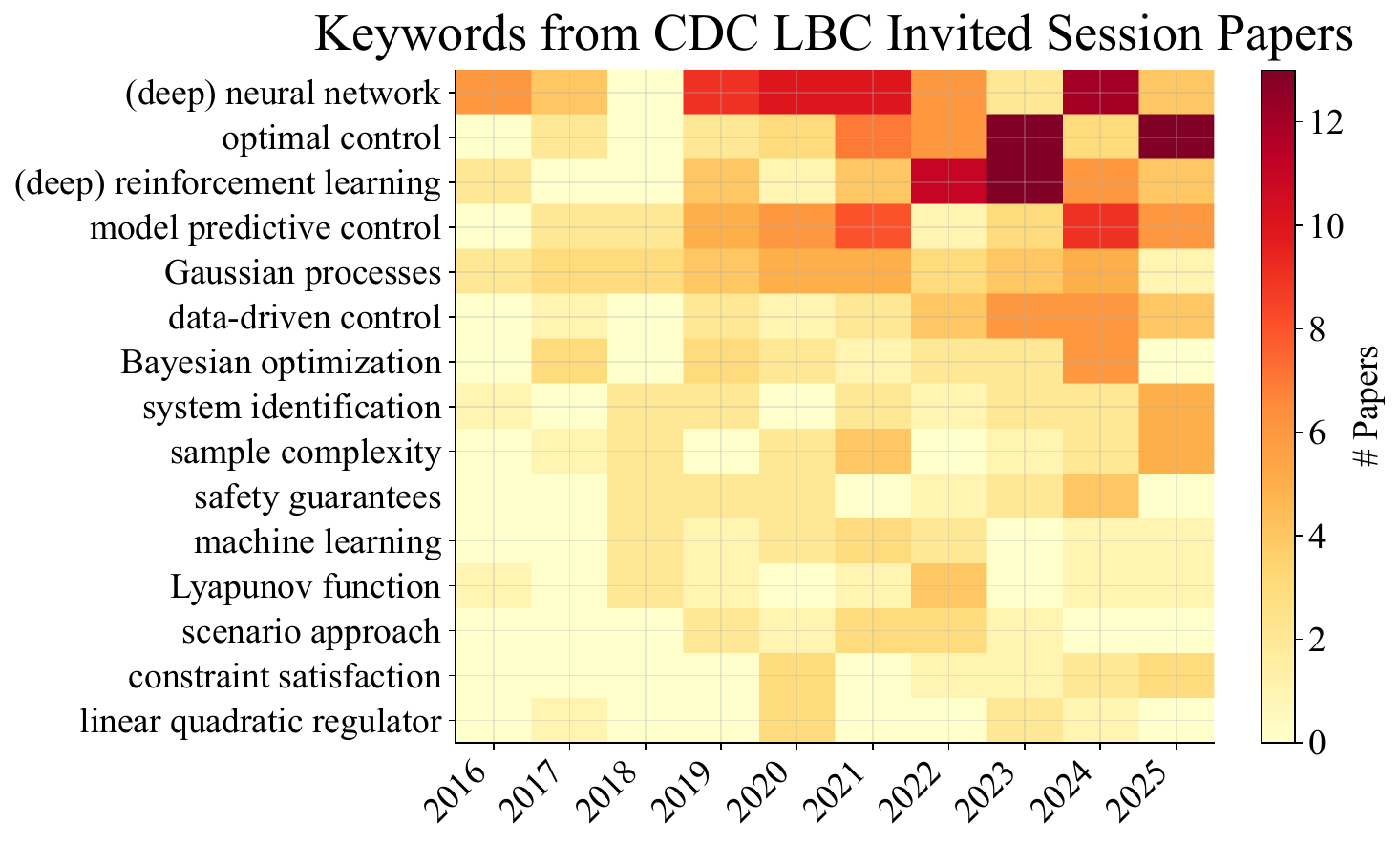}
    \caption{Evolution of recurring research threads in the CDC Learning-Based Control Invited Sessions from 2016 through 2025. The heatmap shows the annual frequency of representative concept phrases extracted from paper titles and abstracts. Details of the keyword extraction and analysis procedure are provided in the appendix.}
    \label{fig:keywords}
\end{figure}

\begin{figure}[t]
    \centering
    \includegraphics[width=\columnwidth]{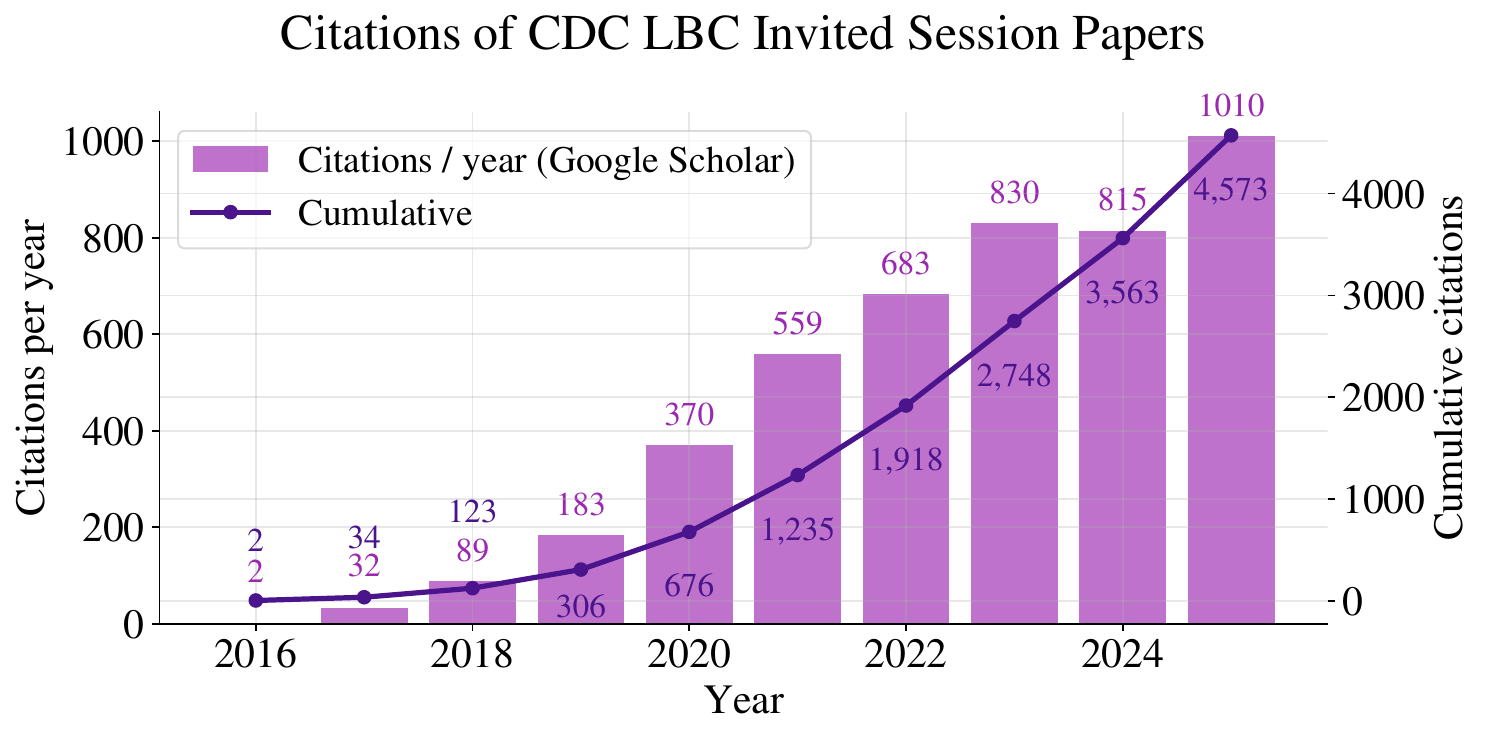}
    \caption{Citation statistics for papers presented in the CDC Learning-Based Control Invited Sessions. The bars show the total number of citations received per year, while the line indicates cumulative citations based on Google Scholar data collected in May 2026.}
    \label{fig:citationstats}
\end{figure}

\textit{Where do these fields converge?}
To foster principled approaches that integrate machine learning and control theory beyond the standard tools of either discipline alone, the Invited Session Series on Learning-Based Control has been held at the IEEE Conference on Decision and Control (CDC) since 2016\footnote{The website of the CDC Learning-Based Control Invited Sessions is available at \href{https://ics.mavt.ethz.ch/Events/learning-control-session.html}{https://ics.mavt.ethz.ch/Events/learning-control-session.html}.\label{footnote:website}}. Over the past decade, this series has hosted 30 dedicated sessions featuring 182 accepted papers, drawing contributions from over 100 distinct institutions. As reflected in~\autoref{fig:keywords}, the series captures the increasing incorporation of expressive learning paradigms into control, while maintaining a strong emphasis on principled methods with desired closed-loop properties. Recurring themes include (deep) neural networks, (deep) \ac{RL}, and \acp{GP}, combined with control tools such as \ac{MPC}, \ac{HJ} reachability, \acp{CLF}, and \acp{CBF} to address performance, stability, and safety. The steady growth in citations over the past decade (\autoref{fig:citationstats}) further reflects the sustained interest in these topics and the expanding body of work contributed by the community.

\begin{figure}[t]
    \centering
    \includegraphics[width=\columnwidth]{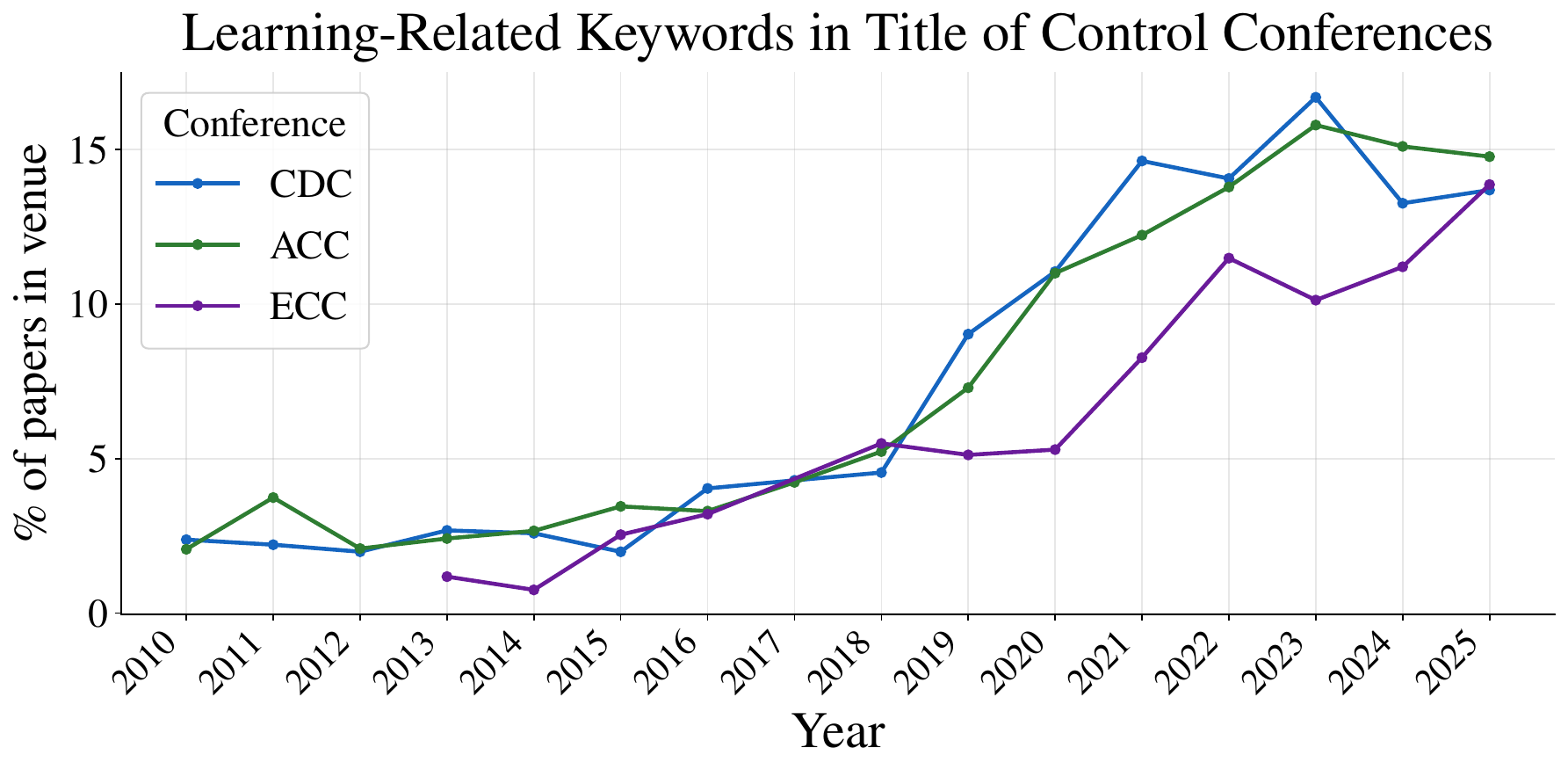}
    \caption{The percentage of papers with the term ``learning'' in their titles across three control conferences (CDC, ACC, and ECC) from 2010 to 2025. Pertinent data was obtained from the DBLP database in May 2026.}
    \label{fig:learningpaperstats}
\end{figure}

This ten-year milestone of the CDC Learning-Based Control Invited Sessions does not exist in isolation. The broader effort to bridge control and learning for safe and scalable autonomy is also reflected in dedicated conferences such as the Learning for Dynamics and Control Conference, as well as in the increasing presence of learning-based methods in core control conferences, as shown in~\autoref{fig:learningpaperstats}.

In this tutorial paper, we highlight key developments and emerging trends in learning-based control over the past decade, spanning control-theoretic foundations, dynamics learning, direct data-driven control, and learning-oriented policy optimization~(see \autoref{fig:overview}). \edits{Our intention is to draw connections among these different families of methods by highlighting how they address uncertainty, which theoretical tools are compatible with them, and what types of guarantees can be established (see~\autoref{tab:method-comparison}).
Rather than providing} an exhaustive survey, \edits{we focus} on representative methodological directions motivated by the CDC Learning-Based Control Invited Sessions. \edits{With the discussions and examples along the way, we hope this tutorial paper serves to} contextualize the progress of the past decade and illustrate how learning and control can be combined to enable more capable, reliable, and safe autonomous systems. 

The paper is organized as follows. We begin by formulating the general learning-based control problem~(\autoref{sec:problem_formulation}). We then review control-theoretic foundations for certifying desired system properties and guarantees~(\autoref{sec:safety}), followed by learning-based control approaches that learn dynamics and improve performance under uncertainty~(\autoref{sec:dynamics}). We next discuss direct data-driven control and learning-oriented methodologies, including \ac{RL}, \ac{IL}, and \ac{BO}, that synthesize or optimize policies from data while preserving desired closed-loop properties~(\autoref{sec:ddc} and~\autoref{sec:rl}). We conclude by highlighting representative applications at the intersection of learning and control~(\autoref{sec:applicationsandbenchmark}), followed by takeaways and future research directions~(\autoref{sec:conclusionsandoutlook}). 

\begin{table*}[t]
\centering
\caption{\centering An overview of the learning-based control method families reviewed in \autoref{sec:dynamics}-\autoref{sec:rl}. For each method family, we detail the specific learning objective, requisite prior knowledge, and the data-driven learning mechanism. Additionally, we summarize how each method quantifies estimation errors, and how those uncertainty models pair with compatible controller synthesis techniques to yield closed-loop guarantees discussed in~\autoref{sec:safety}. \edits{Note that the control methods and closed-loop guarantees presented in the table serve as representative examples, and are as such non-exhaustive. }
}
\label{tab:method-comparison}
\resizebox{\textwidth}{!}{%
\begin{tabular}{@{}|C{2.9cm}|C{3.2cm}|C{2.9cm}|C{3.0cm}|C{2.9cm}|C{3.5cm}|C{3.7cm}|@{}}
\cmidrule[\heavyrulewidth]{1-7}
\textbf{Method Family} & \textbf{What is Learned?} & \textbf{Prior Knowledge Required} & \textbf{Learning from Data} & \textbf{Estimation Errors} & \textbf{Compatible Controller Synthesis Methods} & \textbf{Resulting Closed-Loop Guarantees} \\
\cmidrule[\heavyrulewidth]{1-7}
\multicolumn{7}{@{}l}{\textbf{\textit{Dynamics learning for control} (\autoref{sec:dynamics})}. \textit{All methods can additionally include a known nominal model.} } \\
\cmidrule[\heavyrulewidth]{1-7}

Linear regression (\autoref{subsec:linear})
&
point estimates of coefficients in a linear / fixed-basis nonlinear dynamics model
&
(locally) linear structure /
(nonlinear) basis functions 
&
closed-form least-squares solution based on (offline) input-state data
&
norm bounds or ellipsoidal uncertainty sets for model coefficients 
&
certainty-equivalent and robust LQR;  MPC
&
robust stability; robust constraint satisfaction; probabilistic constraint satisfaction\\
\addlinespace

Neural networks (\autoref{subsec:nn})
&
point estimates of parameters in a nonlinear neural-network dynamics model
&
network architecture and activation functions; optional physical or dynamical structure (e.g., Lipschitz regularities)
&
gradient-based training based on (offline) input-state data
&
deterministic error or output bounds based on regularities
; possibly,  predictive uncertainty estimates via approximate (Bayesian) inference
&
nonlinear \acs{MPC}
;
Lyapunov and small-gain design
&
stability via Lipschitz or local sector bounds; 
\acs{ISS}/\acs{dISS} of recurrent models
\\
\addlinespace

Bayesian linear regression 
(\autoref{subsec:probabilistic}, \autoref{ex:model-learning})
&
posterior distribution over coefficients in a fixed-basis nonlinear dynamics model
&
(nonlinear) basis functions, prior parameter distribution, and noise model
&
closed-form posterior inference based on (offline) input-state data under Gaussian prior and \acs{iid} noise assumptions
&
 probabilistic predictive uncertainty bounds derived from the posterior parameter distribution
&
chance-constrained \acs{MPC} with constraint tightening;  robust MPC with conservative error bounds
&
probabilistic constraint satisfaction or worst-case bounds; 
overconfident if the basis is misspecified
\\
\addlinespace

Gaussian processes  (\autoref{subsec:probabilistic}, \autoref{ex:model-learning}, \autoref{ex:gp-mpc})
&
posterior distribution over the dynamics function represented by a nonparametric model
&
prior mean function, covariance kernel~\cite{pmlr-v28-duvenaud13}, and noise model
&
closed-form posterior inference based on (offline) input-state data under Gaussian prior and \acs{iid}  noise assumptions
&
probabilistic predictive uncertainty bounds derived from the posterior distribution over the dynamics function
&
chance-constrained MPC; 
robust MPC using confidence regions;  
\ac{CLF}-based synthesis
&
probabilistic constraint satisfaction; 
stability via robust invariant terminal sets and worst-case safety from confidence regions
\\
\cmidrule[\heavyrulewidth]{1-7}
\multicolumn{7}{@{}l}{\textit{\textbf{Direct data-driven control} \textbf{(\autoref{sec:ddc})}}} \\
\cmidrule[\heavyrulewidth]{1-7}

Data-based LMI design
&
feedback gain computed from input/state or input/output data
&
LTI system; 
set of basis functions in the nonlinear case
&
rank condition on the collected offline data
&
noise bounds can be used for robust control design
&
state- or output-feedback controllers (including LQR, robust control, etc.)
&
(robust) stability; computable regions of attraction
\\
\addlinespace

Data-driven predictive control (\autoref{ex:deepc})
&
data-based system representation (Willems' lemma)
&
sufficiently long horizon~$L$; suitable regularization tuning
&
rank condition on data; potentially online update of data
&
prediction error bounds depending on noise magnitude
&
linear (and nonlinear) MPC
&
closed-loop stability and robustness; constraint satisfaction
\\
\cmidrule[\heavyrulewidth]{1-7}
\multicolumn{7}{@{}l}{\textit{\textbf{Direct policy learning and controller tuning (\autoref{sec:rl})}}} \\
\cmidrule[\heavyrulewidth]{1-7}

Model-free \acs{RL} (\autoref{ex:rl})
&
policy $\pi_\theta$ and/or action-value function $Q$
&
generally little: reward function and state/action definition, no knowledge of the dynamics
&
 interaction with a simulator or the system (data hungry); offline \acs{RL} from a fixed dataset
&
not modelled explicitly; stochastic policies
&
policy-gradient and actor-critic updates
&
none intrinsic
\\
\addlinespace

Model-based \acs{RL}
&
dynamics model $\hat p_{\theta_M}$, policy, and/or action-value function $Q$
&
same as model-free RL; optional structure in the model
&
 additional interaction with the 
 learned model;
 typically, fewer samples than model-free \acs{RL}
&
uncertainty quantification of learned dynamics model, e.g., via probabilistic ensembles
&
Dyna-style rollouts, or decision-time planning akin to \acs{MPC}
&
safety via Lyapunov functions, 
barrier functions, 
safety filters, 
or \acs{CMDP} soft constraints
\\
\addlinespace

Imitation learning and approximate \acs{MPC}
&
policy $\pi_\theta$ from demonstrations (behaviour cloning) or a reward (inverse \acs{RL})
&
an expert: human demonstrations or an \acs{MPC} to be approximated
&
dataset of state--action pairs, generated from expert; supervised learning
&
approximation error of the learned policy
&
supervised fitting of the \acs{MPC} law; 
constrained generative policies
&
constraint satisfaction despite approximation error via robustness margins or projection
\\
\addlinespace

\acs{BO} controller tuning (\autoref{ex:bo})
&
low-dimensional controller parameters $\theta$ through a surrogate of $J(\theta)$
&
controller structure (e.g., PID, \acs{LQR}, \acs{MPC}); \acs{GP} prior on $J$
&
few closed-loop evaluations, directly on hardware
&
\acs{GP} posterior over the performance $J$
&
applicable to tuning of essentially any controller,
acquisition-function maximization
&
properties inherited from the controller structure; safe tuning under performance and other constraints
\\
\cmidrule[\heavyrulewidth]{1-7}
\end{tabular}%
}
\end{table*}

 The methods and perspectives discussed in this tutorial are drawn from diverse subfields across the machine learning and control communities. For a deeper dive into specific subareas, we refer readers to dedicated surveys, including  learning-based \ac{MPC}~\cite{hewing2020learning}, model-based and safe \ac{RL}~\cite{m2023model,garcia2015comprehensive}, direct data-driven control~\cite{MarkovskyDor2021}, \ac{BO} for controller tuning and robot learning~\cite{stenger2026decade}, \ac{HJ} reachability~\cite{bansal2017hamilton}, \ac{CBF} certification~\cite{ames2019control}, safety filters~\cite{wabersich2023data, hsu_safety_2024}, and safe learning in robotics~\cite{safelearn}. We hope that this tutorial helps contextualize the progress of the past decade and provides a foundation for addressing the challenges that arise as learning-based control continues to scale toward increasingly complex real-world systems.

\section{Problem Formulation}
\label{sec:problem_formulation}
Learning-based control seeks to address general nonlinear control design problems; different families of methods may differ in the specifics of the control problem they address. In the following, we provide a general problem formulation, specializing it in subsequent sections where necessary.

We consider a general, nonlinear dynamical system
\begin{subequations}\label{eq:system_model}
\begin{align}
    x_{k+1} &= f(x_k, u_k, w_k), \label{eq:system} \\
    y_k     &= h(x_k, v_k), \label{eq:measurement}
\end{align}
\end{subequations}
where $k \in \mathbb{Z}_{\ge 0}$ is the discrete-time index, 
$x_k \in \mathbb{R}^{n_\mathrm{x}}$ is the system state, 
$u_k \in \mathbb{R}^{n_\mathrm{u}}$ is the control input, 
$y_k \in \mathbb{R}^{n_\mathrm{y}}$ is the measured output, 
$w_k \in \mathbb{R}^{n_\mathrm{w}}$ is process noise or disturbance, and $v_k \in \mathbb{R}^{n_\mathrm{v}}$ is measurement noise.  The dynamics function~$f$ describes the evolution of the state over time, while $h$ describes how sensor measurements are obtained. Depending on the setting, different specializations of~\eqref{eq:system_model} are considered. 
Common configurations include state feedback (i.e., $y_k = x_k$), 
stochastic noise (i.e., $w_k \sim P_\mathrm{w}$ and $v_k \sim P_\mathrm{v}$ with $P_\mathrm{w}$ and $P_\mathrm{v}$ being appropriately defined distributions), and bounded 
uncertainties (i.e., $w_k \in \mathcal{W}$ and $v_k \in \mathcal{V}$ with $\mathcal{W}\subset \mathbb{R}^{n_\mathrm{w}}$ and $\mathcal{V}\subset \mathbb{R}^{n_\mathrm{v}}$ being bounded sets). While some 
methods specialize to linear dynamics, the strengths of modern 
machine learning typically unfold in the nonlinear setting. The discrete-time 
formulation~\eqref{eq:system_model} is 
predominant in learning-based control and adopted throughout; some methods instead build on continuous-time descriptions, where~\eqref{eq:system} is replaced by a differential equation.

To control the system, we seek a \emph{feedback controller}, also called 
a \textit{policy},
\begin{equation}
    u_k = \pi_k(x_k, r_k) \quad \text{or} \quad 
    u_k = \pi_k(y_k, r_k), \label{eq:policy}
\end{equation}
which maps the current state $x_k$ (or output $y_k$) and possibly a reference signal $r_k$ to the control input $u_k$. The subscript $k$ allows for time-varying policies; stationary policies (i.e., $\pi_k \equiv \pi$) arise as a common special case.

The control objective is often cast as an optimization problem,
\begin{equation}
    \min_{\pi} \; J, \label{eq:objective}
\end{equation}
where the objective function $J$ encodes the desired task such as reference tracking, energy minimization, or disturbance attenuation. It commonly depends on states, inputs, and the reference signal, with its specific arguments depending on the problem. The system may additionally be subject to state and input constraints,
\begin{equation}
    g(x_k, u_k) \leq 0, 
    \label{eq:constraints}
\end{equation}
where $g: \mathbb{R}^{n_\mathrm{x}}\times\mathbb{R}^{n_\mathrm{u}}\to\mathbb{R}^{n_\mathrm{g}}$ denotes the constraint function.

In practice, the system model~\eqref{eq:system_model}, the objective~\eqref{eq:objective}, and the constraints~\eqref{eq:constraints}  may be only partially known or entirely unknown. To address these unknowns, learning-based control leverages machine learning from empirical data, possibly in combination with partial a priori knowledge, to synthesize a policy~\eqref{eq:policy} that reliably achieves the desired objective while satisfying desired safety constraints. The empirical data used in learning-based control often consists of input, state, and/or output measurements:
\begin{subequations}
\label{eq:data_uxy}
\begin{align}
	U & = \begin{bmatrix} u_0^d & u_1^d & \cdots & u_{N-1}^d \end{bmatrix} 
	\in \mathbb{R}^{n_\mathrm{u} \times N}, \label{eq:data_u}\\
    X & = \begin{bmatrix} x_0^d & x_1^d & \cdots & x_{N-1}^d \end{bmatrix} 
	\in \mathbb{R}^{n_\mathrm{x} \times N}, \label{eq:data_x} \\
    Y & = \begin{bmatrix} y_0^d & y_1^d & \cdots & y_{N-1}^d \end{bmatrix} 
    \in \mathbb{R}^{n_\mathrm{y} \times N}, \label{eq:data_y}
\end{align} 
\end{subequations}
where $N$ denotes the length of the data sequence and the superscript~``$d$'' denotes the (pre-)recorded dataset. For convenience, we further define $X^+$ as the state data shifted by one time step:
\begin{equation}
    X^+ = \begin{bmatrix} x_1^d & x_2^d & \cdots & x_{N}^d \end{bmatrix}
    \in \mathbb{R}^{n_\mathrm{x} \times N}. \label{eq:data_x+}
\end{equation}
Many problems in learning-based control assume that state measurements are available and use input and state data, i.e.,~\eqref{eq:data_u} and \eqref{eq:data_x}, for learning. 
However, particularly challenging learning problems also arise when only output measurements are available, i.e.,~\eqref{eq:data_u} and~\eqref{eq:data_y}. Such problems often give rise to (learning-based) combinations of state estimation and 
system identification~\cite{masti2018learning, doerr2018probabilistic, gedon2021deep, buisson2021joint, bin2021model, barfoot2024state, pillonetto2025deep}, which also relate to latent-space modeling in machine learning.

While the focus of this tutorial paper is on learning with state feedback and learning-based \emph{control}, problems based on output measurements and state estimation are discussed  at selected points throughout. Likewise, while we focus on \emph{offline} learning, where the phase of  data collection and learning is typically separate from controller operation,  the connection to online learning, i.e., updating learned models or policies during operation, is  discussed at relevant places.
\section{Stability and Safety}
~\label{sec:safety}
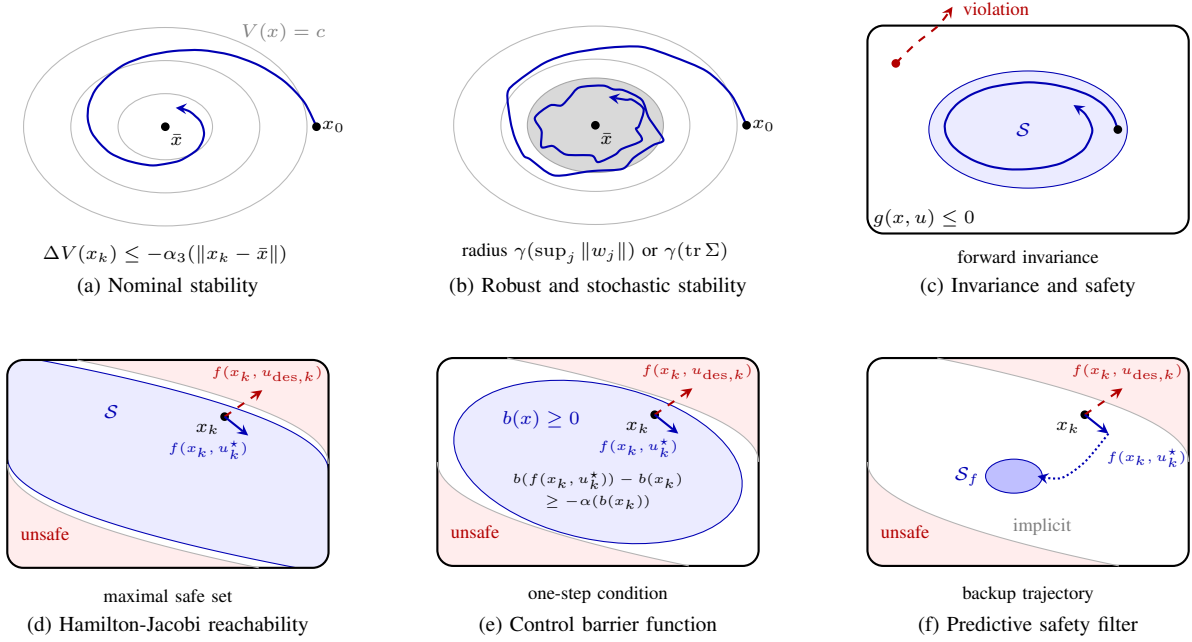
\begin{figure*}[t]
	\ifdefined\variantname\centering{\bfseries\variantname\par}\medskip\fi  
	\centering
	\begin{subfigure}[b]{0.32\textwidth}
		\centering
		\begin{tikzpicture}[scale=1.25, >=stealth]
			\foreach \r in {0.5,1.0,1.5} {\draw[gray!55, thin] (0,0) ellipse ({\r} and {0.7*\r});}
			\node[gray, font=\scriptsize] at (1.25,1.0) {$V(x)=c$};
			\draw[thick, blue!70!black, ->] plot[smooth] coordinates {(1.60,0.00) (1.47,0.26) (1.26,0.48) (0.99,0.65) (0.69,0.76) (0.38,0.81) (0.08,0.80) (-0.19,0.75) (-0.43,0.66) (-0.61,0.53) (-0.74,0.39) (-0.81,0.23) (-0.82,0.08) (-0.78,-0.06) (-0.69,-0.18) (-0.58,-0.28) (-0.43,-0.35) (-0.28,-0.39) (-0.13,-0.41) (0.02,-0.39) (0.15,-0.36) (0.26,-0.30) (0.34,-0.24) (0.39,-0.16) (0.41,-0.08) (0.40,-0.01) (0.37,0.06) (0.32,0.11) (0.26,0.16) (0.18,0.19) (0.11,0.20)};
			\fill (0,0) circle (1.3pt);
			\node[font=\scriptsize, below right, inner sep=1pt] at (0.02,-0.05) {$\bar x$};
			\fill (1.6,0) circle (1.3pt);
			\node[font=\scriptsize, right, inner sep=2pt] at (1.6,0) {$x_0$};
			\node[font=\scriptsize] at (0,-1.35) {$\Delta V(x_k) \le -\alpha_3(\|x_k-\bar x\|)$};
			\path (-1.9,-1.5) rectangle (1.9,1.5);
		\end{tikzpicture}
		\caption{\edits{Nominal stability}}
		\label{fig:ss_nominal}
	\end{subfigure}%
	\begin{subfigure}[b]{0.32\textwidth}
		\centering
		\begin{tikzpicture}[scale=1.25, >=stealth]
			\foreach \r in {1.0,1.5} {\draw[gray!55, thin] (0,0) ellipse ({\r} and {0.7*\r});}
			\fill[gray!30] (0,0) ellipse (0.72 and 0.5);
			\draw[gray!70, thin] (0,0) ellipse (0.72 and 0.5);
			\draw[thick, blue!70!black, ->] plot[smooth] coordinates {(1.60,0.00) (1.44,0.32) (1.18,0.57) (0.81,0.73) (0.49,0.83) (0.12,0.82) (-0.26,0.76) (-0.54,0.66) (-0.83,0.46) (-0.93,0.34) (-0.90,0.05) (-0.85,-0.14) (-0.76,-0.32) (-0.69,-0.36) (-0.44,-0.53) (-0.09,-0.49) (0.03,-0.48) (0.29,-0.46) (0.41,-0.34) (0.61,-0.21) (0.72,-0.16) (0.64,-0.03) (0.56,0.09) (0.49,0.27) (0.30,0.37) (0.19,0.38) (0.08,0.42) (-0.17,0.40) (-0.27,0.30) (-0.36,0.21) (-0.48,0.21) (-0.64,0.08) (-0.57,-0.06) (-0.57,-0.23) (-0.45,-0.22) (-0.31,-0.31) (-0.04,-0.32) (0.03,-0.39) (0.21,-0.31) (0.28,-0.25) (0.49,-0.16) (0.43,-0.04) (0.46,-0.01) (0.51,0.17) (0.39,0.27) (0.32,0.27) (0.14,0.30)};
			\fill (0,0) circle (1.3pt);
			\node[font=\scriptsize, below right, inner sep=1pt] at (0.02,-0.05) {$\bar x$};
			\fill (1.6,0) circle (1.3pt);
			\node[font=\scriptsize, right, inner sep=2pt] at (1.6,0) {$x_0$};
			\node[font=\scriptsize] at (0,-1.35) {radius $\gamma(\sup_j\|w_j\|)$ or $\gamma(\operatorname{tr}\Sigma)$};
			\path (-1.9,-1.5) rectangle (1.9,1.5);
		\end{tikzpicture}
		\caption{\edits{Robust and stochastic stability}}
		\label{fig:ss_robust}
	\end{subfigure}%
	\begin{subfigure}[b]{0.32\textwidth}
		\centering
		\begin{tikzpicture}[scale=1.25, >=stealth]
			\draw[thick, rounded corners=6pt] (-1.7,-1.1) rectangle (1.7,1.1);
			\node[font=\scriptsize, above right, inner sep=2pt] at (-1.7,-1.1) {$g(x,u)\le 0$};
			\fill[blue!8] (0,0) ellipse (1.05 and 0.62);
			\draw[blue!70!black, thin] (0,0) ellipse (1.05 and 0.62);
			\node[font=\scriptsize, blue!70!black] at (-0.05,0) {$\mathcal S$};
			\draw[thick, blue!70!black, ->] plot[smooth] coordinates {(0.95,0.00) (0.90,0.13) (0.80,0.26) (0.65,0.36) (0.46,0.43) (0.24,0.48) (0.01,0.50) (-0.21,0.48) (-0.42,0.44) (-0.60,0.37) (-0.74,0.28) (-0.83,0.18) (-0.88,0.06) (-0.87,-0.05) (-0.81,-0.16) (-0.71,-0.25) (-0.57,-0.33) (-0.40,-0.39) (-0.21,-0.42) (-0.02,-0.42) (0.17,-0.40) (0.34,-0.36) (0.48,-0.29) (0.59,-0.21) (0.65,-0.11) (0.68,-0.01) (0.65,0.09) (0.59,0.18) (0.48,0.26)};
			\fill (0.95,0) circle (1.3pt);
			\fill[red!70!black] (-1.4,0.7) circle (1.3pt);
			\draw[thick, red!70!black, dashed, ->] plot[smooth] coordinates {(-1.4,0.7) (-1.2,0.9) (-0.95,1.1) (-0.8,1.3)};
			\node[font=\scriptsize, red!70!black, right, inner sep=2pt] at (-0.75,1.28) {violation};
			\node[font=\scriptsize] at (0,-1.35) {forward invariance};
			\path (-1.9,-1.5) rectangle (1.9,1.5);
		\end{tikzpicture}
		\caption{\edits{Invariance and safety}}
		\label{fig:ss_invariance}
	\end{subfigure}\\[0.8em]
	\begin{subfigure}[b]{0.32\textwidth}
		\centering
		\begin{tikzpicture}[scale=1.25, >=stealth]
			\def\sfmargin{0.08}
			\begin{scope}
				\clip[rounded corners=6pt] (-1.7,-1.1) rectangle (1.7,1.1);
				\fill[red!7] (1.7,1.1) -- plot[variable=\t, domain=0:1.1, samples=30] ({1.7-2.2*(1.1-\t)*(1.1-\t)},{1.1-\t}) -- cycle;
				\fill[red!7] (-1.7,-1.1) -- plot[variable=\t, domain=0:1.1, samples=30] ({-1.7+2.2*(1.1-\t)*(1.1-\t)},{-1.1+\t}) -- cycle;
				\fill[blue!8] (-1.7,1.1) -- plot[variable=\t, domain=0:1.1+\sfmargin, samples=40] ({1.7-2.2*(1.1+\sfmargin-\t)*(1.1+\sfmargin-\t)},{1.1-\t}) -- (1.7,-1.1) -- plot[variable=\t, domain=0:1.1+\sfmargin, samples=40] ({-1.7+2.2*(1.1+\sfmargin-\t)*(1.1+\sfmargin-\t)},{-1.1+\t}) -- cycle;
				\draw[blue!70!black, thin] plot[variable=\t, domain=0:1.1+\sfmargin, samples=40] ({1.7-2.2*(1.1+\sfmargin-\t)*(1.1+\sfmargin-\t)},{1.1-\t});
				\draw[blue!70!black, thin] plot[variable=\t, domain=0:1.1+\sfmargin, samples=40] ({-1.7+2.2*(1.1+\sfmargin-\t)*(1.1+\sfmargin-\t)},{-1.1+\t});
			\end{scope}
			\draw[thick, rounded corners=6pt] (-1.7,-1.1) rectangle (1.7,1.1);
			\draw[gray!60, thin] plot[variable=\t, domain=0:1.1, samples=30] ({1.7-2.2*(1.1-\t)*(1.1-\t)},{1.1-\t});
			\draw[gray!60, thin] plot[variable=\t, domain=0:1.1, samples=30] ({-1.7+2.2*(1.1-\t)*(1.1-\t)},{-1.1+\t});
			\node[font=\scriptsize, blue!70!black] at (-0.6,0.55) {$\mathcal S$};
			\node[font=\scriptsize, red!70!black] at (-1.3,-0.74) {unsafe};
			\node[font=\scriptsize] at (0,-1.4) {maximal safe set};
			\fill (0.6,0.5) circle (1.3pt);
			\node[font=\scriptsize, below left, inner sep=1pt] at (0.58,0.46) {$x_k$};
			\draw[thick, red!70!black, dashed, ->] (0.6,0.5) -- (1.0,0.78);
			\node[font=\tiny, red!70!black, above, inner sep=1pt] at (1.05,0.82) {$f(x_k,u_{\mathrm{des},k})$};
			\draw[thick, blue!70!black, ->] (0.6,0.5) -- (0.85,0.3);
			\node[font=\tiny, blue!70!black, below left, inner sep=1pt] at (0.88,0.28) {$f(x_k,u_k^\star)$};
			\path (-1.9,-1.5) rectangle (1.9,1.5);
		\end{tikzpicture}
		\caption{\edits{Hamilton-Jacobi reachability}}
		\label{fig:ss_hj}
	\end{subfigure}%
	\begin{subfigure}[b]{0.32\textwidth}
		\centering
		\begin{tikzpicture}[scale=1.25, >=stealth]
			\begin{scope}
				\clip[rounded corners=6pt] (-1.7,-1.1) rectangle (1.7,1.1);
				\fill[red!7] (1.7,1.1) -- plot[variable=\t, domain=0:1.1, samples=30] ({1.7-2.2*(1.1-\t)*(1.1-\t)},{1.1-\t}) -- cycle;
				\fill[red!7] (-1.7,-1.1) -- plot[variable=\t, domain=0:1.1, samples=30] ({-1.7+2.2*(1.1-\t)*(1.1-\t)},{-1.1+\t}) -- cycle;
			\end{scope}
			\draw[thick, rounded corners=6pt] (-1.7,-1.1) rectangle (1.7,1.1);
			\draw[gray!60, thin] plot[variable=\t, domain=0:1.1, samples=30] ({1.7-2.2*(1.1-\t)*(1.1-\t)},{1.1-\t});
			\draw[gray!60, thin] plot[variable=\t, domain=0:1.1, samples=30] ({-1.7+2.2*(1.1-\t)*(1.1-\t)},{-1.1+\t});
			\node[font=\scriptsize, red!70!black] at (-1.3,-0.74) {unsafe};
			\fill[blue!8, rotate=-12] (0,0) ellipse (1.55 and 0.82);
			\draw[blue!70!black, thin, rotate=-12] (0,0) ellipse (1.55 and 0.82);
			\node[font=\scriptsize, blue!70!black] at (-0.6,0.45) {$b(x)\ge0$};
			\node[font=\tiny, align=center] at (0.0,-0.3) {$b(f(x_k,u_k^\star)) - b(x_k)$\\[2pt]$\ge -\alpha(b(x_k))$};
			\node[font=\scriptsize] at (0,-1.4) {one-step condition};
			\fill (0.6,0.5) circle (1.3pt);
			\node[font=\scriptsize, below left, inner sep=1pt] at (0.58,0.46) {$x_k$};
			\draw[thick, red!70!black, dashed, ->] (0.6,0.5) -- (1.0,0.78);
			\node[font=\tiny, red!70!black, above, inner sep=1pt] at (1.05,0.82) {$f(x_k,u_{\mathrm{des},k})$};
			\draw[thick, blue!70!black, ->] (0.6,0.5) -- (0.85,0.3);
			\node[font=\tiny, blue!70!black, below left, inner sep=1pt] at (0.88,0.28) {$f(x_k,u_k^\star)$};
			\path (-1.9,-1.5) rectangle (1.9,1.5);
		\end{tikzpicture}
		\caption{\edits{Control barrier function}}
		\label{fig:ss_cbf}
	\end{subfigure}%
	\begin{subfigure}[b]{0.32\textwidth}
		\centering
		\begin{tikzpicture}[scale=1.25, >=stealth]
			\begin{scope}
				\clip[rounded corners=6pt] (-1.7,-1.1) rectangle (1.7,1.1);
				\fill[red!7] (1.7,1.1) -- plot[variable=\t, domain=0:1.1, samples=30] ({1.7-2.2*(1.1-\t)*(1.1-\t)},{1.1-\t}) -- cycle;
				\fill[red!7] (-1.7,-1.1) -- plot[variable=\t, domain=0:1.1, samples=30] ({-1.7+2.2*(1.1-\t)*(1.1-\t)},{-1.1+\t}) -- cycle;
			\end{scope}
			\draw[thick, rounded corners=6pt] (-1.7,-1.1) rectangle (1.7,1.1);
			\draw[gray!60, thin] plot[variable=\t, domain=0:1.1, samples=30] ({1.7-2.2*(1.1-\t)*(1.1-\t)},{1.1-\t});
			\draw[gray!60, thin] plot[variable=\t, domain=0:1.1, samples=30] ({-1.7+2.2*(1.1-\t)*(1.1-\t)},{-1.1+\t});
			\node[font=\scriptsize, red!70!black] at (-1.3,-0.74) {unsafe};
			\node[font=\scriptsize, gray] at (0.15,-0.68) {implicit};
			\fill[blue!25] (-0.15,-0.15) ellipse (0.3 and 0.18);
			\draw[blue!70!black, thin] (-0.15,-0.15) ellipse (0.3 and 0.18);
			\node[font=\scriptsize, blue!70!black, left, inner sep=2pt] at (-0.45,-0.15) {$\mathcal S_f$};
			\node[font=\scriptsize] at (0,-1.4) {backup trajectory};
			\fill (0.6,0.5) circle (1.3pt);
			\node[font=\scriptsize, below left, inner sep=1pt] at (0.58,0.46) {$x_k$};
			\draw[thick, red!70!black, dashed, ->] (0.6,0.5) -- (1.0,0.78);
			\node[font=\tiny, red!70!black, above, inner sep=1pt] at (1.05,0.82) {$f(x_k,u_{\mathrm{des},k})$};
			\draw[thick, blue!70!black, ->] (0.6,0.5) -- (0.85,0.3);
			\node[font=\tiny, blue!70!black, right, inner sep=1pt] at (0.8,0.04) {$f(x_k,u_k^\star)$};
			\draw[blue!70!black, densely dotted, thick, ->] plot[smooth] coordinates {(0.85,0.3) (0.6,0.0) (0.35,-0.17) (0.1,-0.15)};
			\path (-1.9,-1.5) rectangle (1.9,1.5);
		\end{tikzpicture}
		\caption{\edits{Predictive safety filter}}
		\label{fig:ss_psf}
	\end{subfigure}
	\caption{\edits{Illustration of the stability and safety notions of this section. (a) Nominal stability: a Lyapunov function $V$ decreases along closed-loop trajectories (level sets in grey) and the state converges to the equilibrium $\bar x$. (b) Robust and stochastic stability: under disturbances $w$, the state converges to a neighbourhood of $\bar x$ (grey), whose size depends on the disturbance bound or the noise covariance. (c) Safety: constraint satisfaction at one time instant does not imply safety (red dashed arrow); safety is guaranteed by an invariant set $\mathcal S$ (blue) contained in the constraint set (white). (d)--(f) Safety filters: the desired input $u_{\mathrm{des},k}$ would violate the constraint and is  modified to $u_k^\star$ to stay inside of the constraint set (white). HJ reachability computes the maximal safe set $\mathcal S$ offline, the largest invariant set within the constraint set. A CBF certifies invariance of $\{b(x)\ge0\}$ by the one-step condition on $b(x_{k+1})$. A PSF does not compute the safe set but certifies safety by a backup trajectory (dotted) into a smaller terminal set $\mathcal S_f$. }}
	\label{fig:stability_safety}
\end{figure*}

We begin by introducing the control-theoretic foundations that provide the desired closed-loop properties of stability and constraint satisfaction. We first review these concepts in the nominal setting, in which the dynamics are known exactly, and then extend the discussion to robust and stochastic formulations that explicitly account for uncertainty. The latter settings are particularly relevant in learning-based control, where the dynamics are only approximately known from data and the resulting model error must be incorporated into the analysis either as a bounded disturbance or as a probabilistic uncertainty description. \edits{\autoref{fig:stability_safety} illustrates the main concepts of this section.}

We first recall the standard comparison function classes used throughout this section~\cite{khalil2002nonlinear}. A function $\alpha : \mathbb{R}_{\ge 0} \to \mathbb{R}_{\ge 0}$ is of class $\mathcal{K}$ if it is continuous, strictly increasing, and satisfies $\alpha(0)=0$. It is of class $\mathcal{K}_\infty$ if additionally $\alpha(r)\to\infty$ as $r\to\infty$. A function $\beta : \mathbb{R}_{\ge0}\times\mathbb{R}_{\ge0}\to\mathbb{R}_{\ge0}$ is of class $\mathcal{KL}$ if $\beta(\cdot,s)\in\mathcal K$ for each fixed $s$, and $\beta(r,s)\to0$ as $s\to\infty$ for each fixed $r$.

\subsection{\texorpdfstring{\edits{Nominal Stability}}{Nominal Stability}}
Stability concerns the asymptotic behaviour of a system around an equilibrium. Throughout this section, unless otherwise stated, the properties considered are local, i.e., they hold in a neighbourhood of the equilibrium.
When the dynamics $f$ are known, stability concerns an equilibrium $\bar x$ of system~\eqref{eq:system} in closed loop with a policy $u_k=\pi(x_k)$.

\begin{defn}[Nominal stability~\cite{khalil2002nonlinear}]\label{def:lyapunov}
	Let $\bar x$ be an equilibrium of system~\eqref{eq:system} under the
	policy $u_k=\pi(x_k)$ and with $w_k \equiv 0$. The equilibrium $\bar x$ is
	\begin{itemize}
		\item stable if there exist $\gamma \in \mathcal K$ and $r > 0$ such that
		      \[
			      \|x_k - \bar x\|
			      \le
			      \gamma(\|x_0 - \bar x\|)
		      \]
		      for all $ k \ge 0$ and all initial conditions satisfying $\|x_0 - \bar x\| < r$;
		\item asymptotically stable if there exist $\beta\in\mathcal{KL}$ and $r > 0$ such that
		      \[
			      \|x_k-\bar x\|
			      \le
			      \beta(\|x_0-\bar x\|,k)
		      \]
		      for all $k \ge 0$ and all initial conditions satisfying  $\|x_0 - \bar x\| < r$.
	\end{itemize}
\end{defn}

Lyapunov's direct method certifies stability through a pointwise decrease condition on a scalar energy-like function.

\begin{thm}[Lyapunov's direct method~\cite{khalil2002nonlinear}]\label{thm:lyapunov}
	Let $\bar x$ be an equilibrium and $\mathcal{D}$ a neighbourhood of
	$\bar x$. Suppose there exists a function
	$V : \mathcal{D} \to \mathbb{R}$ satisfying
	\[
		\alpha_1(\|x-\bar x\|)
		\le
		V(x)
		\le
		\alpha_2(\|x-\bar x\|)
	\]
	for some $\alpha_1,\alpha_2\in\mathcal K_\infty$, and let
	\[
		\Delta V(x_k)
		:=
		V(x_{k+1})-V(x_k)
	\]
	denote its one-step change along the closed-loop trajectories. Then:
	\begin{enumerate}
		\item if $\Delta V(x_k)\le0$ on $\mathcal D$, then $\bar x$ is stable;
		\item if there exists $\alpha_3\in\mathcal K_\infty$ such that
		      \[
			      \Delta V(x_k)
			      \le
			      -\alpha_3(\|x_k-\bar x\|)
		      \]
		      on $\mathcal D$, then $\bar x$ is asymptotically stable.
	\end{enumerate}
\end{thm}

\subsection{\texorpdfstring{\edits{Robust Stability}}{Robust Stability}}
In practice, however, the dynamics model used for control design is rarely exact. Learning-based controllers rely on models identified from finite data and therefore inevitably incur approximation error. This motivates robust stability notions that explicitly account for uncertainty and disturbances.

In the robust setting, the disturbance is assumed bounded, i.e., $w\in\mathcal{W}$, with $\mathcal{W}\subset \mathbb{R}^{n_\mathrm{w}}$ a compact set.
\edits{In the following, the model error and external disturbances are collected in the disturbance term $w$ of~\eqref{eq:system}. A commonly used robust stability notion is then \ac{ISS}, which characterizes the effect of the disturbance on the state trajectory.}

\begin{defn}[Input-to-state stability~\cite{khalil2002nonlinear}]\label{def:iss}
	Let $\bar x$ be an equilibrium of the system with $w \equiv \{0\}$, and consider
	\eqref{eq:system} in closed loop with $u_k = \pi(x_k)$ and disturbance
	sequence $w = \{w_k\}$, $w_k\in\mathcal{W}$. The closed-loop system is input-to-state stable
	if there exist $\beta \in \mathcal{KL}$ and $\gamma \in \mathcal{K}$
	such that, for every initial state $x_0$ and every bounded disturbance
	sequence~$w$,
	\begin{equation*}
		\|x_k - \bar x\|
		\le
		\beta(\|x_0 - \bar x\|,\, k)
		+
		\gamma\:\:\Big(\sup_{0 \le j < k} \|w_j\|\Big),
		\qquad k \ge 0 .
		\vspace{0.5em}
	\end{equation*}
\end{defn}

The ISS property separates the influence of the initial condition from that of the disturbance. The first term decays with time, whereas the second quantifies the residual error induced by uncertainty. 
As in the nominal case, ISS admits a Lyapunov characterization.

\begin{thm}[ISS-Lyapunov function~\cite{khalil2002nonlinear}]\label{thm:iss}
	Suppose there exists a function $V$ and
	$\alpha_1, \alpha_2 \in \mathcal{K}_\infty$ with
	\[
		\alpha_1(\|x - \bar x\|)
		\le
		V(x)
		\le
		\alpha_2(\|x - \bar x\|),
	\]
	whose one-step change along the closed-loop trajectories satisfies
	\begin{equation*}
		V(x_{k+1}) - V(x_k)
		\le
		-\alpha_3(\|x_k - \bar x\|)
		+
		\sigma(\|w_k\|)
	\end{equation*}
	for some $\alpha_3 \in \mathcal{K}_\infty$ and $\sigma \in \mathcal{K}$.
	Then the closed-loop system is input-to-state stable in the sense of~\autoref{def:iss}.
\end{thm}

When $w = \{0\}$, the ISS property reduces to
\[
	\|x_k - \bar x\|
	\le
	\beta(\|x_0 - \bar x\|, k),
\]
thus recovering the asymptotic stability of the nominal system.

A related robustness notion is incremental stability~\cite{angeli2002lyapunov}, which studies the convergence between pairs of trajectories rather than toward a single equilibrium. \edits{This concept extends to \ac{dISS}. Such properties can be used in order to compare nominal and perturbed trajectories in the presence of model mismatch.}

\subsection{\texorpdfstring{\edits{Stochastic Stability}}{Stochastic Stability}}
When the disturbance $w_k$ in~\eqref{eq:system} is modelled probabilistically, with $w_k \sim P_w$, the closed-loop system becomes a stochastic process. In this setting, stability is commonly formulated either in probability or in expectation.

\begin{defn}[Stochastic stability~\cite{mcallister2023stochastic,chatterjee2015stability}]\label{def:stochstab}
	Let $\bar x$ be an equilibrium for the system with $w \equiv \{0\}$, and consider~\eqref{eq:system} in closed loop with $u_k = \pi(x_k)$ and \ac{iid}, zero-mean disturbance $w_k \sim P_w$ with covariance $\Sigma$ and finite second moment. The equilibrium $\bar x$ is
	\begin{itemize}
		\item mean-square bounded if
		      \[
			      \limsup_{k \to \infty}\:\:
			      \mathbb{E}\,\|x_k - \bar x\|^2 < \infty .
		      \]
		\item robustly asymptotically stable in expectation if there exist $\beta \in \mathcal{KL}$ and $\gamma \in \mathcal{K}$ such that
		      \[
			      \mathbb{E}\,\|x_k - \bar x\|
			      \le \beta(\|x_0 - \bar x\|,\,k) + \gamma(\operatorname{tr}\Sigma),
			      \quad \forall k \geq 0.
		      \]
		\item asymptotically stable in expectation if it is robustly asymptotically stable in expectation with $\gamma \equiv 0$, i.e., there exists  $\beta \in \mathcal{KL}$ such that
		      \[
			      \mathbb{E}\,\|x_k - \bar x\|
			      \le \beta(\|x_0 - \bar x\|,\,k),
			      \quad \forall k \geq 0.
		      \]
	\end{itemize}
\end{defn}

As in the deterministic setting, stochastic stability can be established through a Lyapunov function satisfying an expected decrease condition.

\begin{thm}[Stochastic Lyapunov function~\cite{mcallister2023stochastic,chatterjee2015stability}]\label{thm:stochstab}
	Suppose there exist a function $V$ and
	$\alpha_1, \alpha_2 \in \mathcal{K}_\infty$ with
	\[
		\alpha_1(\|x - \bar x\|)
		\le
		V(x)
		\le
		\alpha_2(\|x - \bar x\|),
	\]
	a positive-definite stage cost $\ell$, and a constant
	$c \ge 0$ such that
	\begin{equation}
    \label{eq:lyap_dec_stoc}
		\mathbb{E}\!\left[V(x_{k+1}) \mid x_k\right] - V(x_k)
		\le
		-\ell(x_k) + c .
	\end{equation}
	Then, the closed loop is mean-square bounded and satisfies
	\begin{equation*}
		\limsup_{N \to \infty}
		\frac{1}{N}
		\sum_{k=0}^{N-1}
		\mathbb{E}\,\ell(x_k)
		\le c ,
	\end{equation*}
	and the equilibrium $\bar x$ is robustly asymptotically stable in expectation.
	If moreover $c = 0$, the equilibrium $\bar x$ is asymptotically stable in expectation.
\end{thm}

The constant $c$ in~\eqref{eq:lyap_dec_stoc} measures the residual effect of stochastic uncertainty and plays a role analogous to the disturbance gain in ISS. When $c = 0$ there is no residual stochastic term and $\bar{x}$ is recovered as an asymptotically stable equilibrium in expectation. Otherwise, the state remains bounded in mean square within a neighbourhood of $\bar x$, while the long-term average performance remains bounded by~$c$.

\subsection{\texorpdfstring{\edits{Invariance and Safety Filters}}{Invariance and Safety Filters}}
While stability concerns convergence toward equilibria, safety concerns the satisfaction of state and input constraints~\eqref{eq:constraints} over time.
\edits{Note that constraint satisfaction at a single time instant does not guarantee constraint satisfaction at future times, since there exist states satisfying the constraints from which a future constraint violation cannot be avoided under the system dynamics.} Safety therefore requires identifying sets from which the constraints remain satisfied for all future times. The central notion underlying safety (in terms of constraint satisfaction) is thus invariance.

\begin{defn}[Forward invariance~\cite{blanchini1999invariance}]
	Consider system~\eqref{eq:system} in closed loop with the policy $u_k=\pi(x_k)$, and let $\mathcal S \subseteq \mathbb R^{n_\mathrm{x}}$. The set $\mathcal S$ is said to be forward invariant if, for all initial conditions $x_0 \in \mathcal S$, the corresponding trajectory satisfies $x_k \in \mathcal S$ for all $k \ge 0$.
\end{defn}

A forward invariant set guarantees that trajectories initialized within it remain inside for all future times; safety then follows whenever such a set lies within the constraints. When the input is a decision variable, the relevant notion is control invariance, i.e., for every $x\in \mathcal S$, the existence of an input $u$ satisfying~\eqref{eq:constraints} and keeping the state in~$\mathcal S$.

Invariant sets can either be integrated directly as constraints in the control problem, as e.g., common practice in \ac{MPC}, \edits{or can be used to define a safety filter.} Safety filters are independent of the control policy and merely verify the safety of a proposed control action, and otherwise implement a safety-ensuring modification. We focus on the safety filter formulation in the following. For a tutorial overview of common safety filter formulations and their data-driven extensions, see also e.g., \cite{wabersich2023data}.

Classical approaches to safety characterize invariant sets through reachability analysis. In particular, \ac{HJ} reachability methods~\cite{bansal2017hamilton} compute the set of states from which trajectories can be kept within the constraints under the system dynamics by propagating backward the set of constraint-violating states. Such methods provide a global characterization of safety and typically yield tight approximations of the safe set under the assumed dynamics. However, their computational complexity scales poorly with the dimension of the state space, which limits their applicability to high-dimensional systems and has motivated various approximate
and learning-based extensions~\cite{wabersich2023data,bansal2017hamilton,mochen2018hamilton}.

An alternative approach relies on \acp{CBF}~\cite{ames2019control}, which enforce safety locally online through tractable conditions on the closed-loop dynamics. Analogously to Lyapunov functions for stability, barrier functions certify safety through conditions that guarantee forward invariance of a safe set. Consider a safe set of the form
\[
	\mathcal S := \{x \in \mathbb R^{n_\mathrm{x}} : b(x)\ge0\},
\]
where $b : \mathbb R^{n_\mathrm{x}} \to \mathbb R$. In discrete time, given a valid \ac{CBF}, one can  impose conditions on the control input such that
\[
	b(f(x_k,u_k)) - b(x_k)
	\ge
	-\alpha(b(x_k))
\]
for some $\alpha \in \mathcal K$ satisfying $\alpha(r) \le r$ for all $r \ge 0$. This condition ensures that trajectories cannot leave the safe set~$\mathcal S$, thereby rendering the set forward invariant. \edits{In discrete time, the barrier condition directly provides a one-step-ahead safety filter.}  
A common optimization problem is given by
\[
	\begin{aligned}
		u_k^\star
		:=
		\arg\min_{u_k}
		\quad              &
		\|u_k-u_{\mathrm{des},k}\|^2 \\
		\mathrm{s.t.}\quad &
		b(f(x_k,u_k)) - b(x_k)
		\ge
		-\alpha(b(x_k)),
	\end{aligned}
\]
where $u_{\mathrm{des},k}$ denotes the desired input generated by the nominal controller or learned policy at time $k$.

\edits{When comparing the computational cost of the two approaches, the construction of the safety certificate has to be distinguished from its online use. HJ reachability computes the value function on a grid over the state space. This is expensive and scales exponentially with the state dimension, but yields the (approximate) maximal safe set. A CBF only requires a function $b$ satisfying the barrier condition, which is cheaper to obtain but generally provides an inner approximation of the safe set \edits{and is in general difficult to construct for nonlinear systems.} Once the certificate is available, both filters are cheap to evaluate online, since the HJ value function can be used in the same way as a barrier function~\cite{borquez2024safety}.}

\edits{\Acp{PSF}~\cite{wabersich2018linear, wabersich2021predictive} do not require the construction of a large safe set, but enforce safety by means of finite-horizon trajectory planning.} Rather than explicitly computing a global invariant set, PSFs certify safety by ensuring the existence of a feasible backup trajectory that steers the system to a local invariant set.
Given the current state and a desired input proposed by a controller or learning-based policy, a \ac{PSF} solves a constrained finite-horizon optimal control problem of the form
\[
	\begin{aligned}
		\min_{\{u_{i|k}\}}
		\quad              &
		\|u_{0|k}-u_{\mathrm{des},k}\|^2                                     \\
		\mathrm{s.t.}\quad &
		x_{0|k}=x_k,                                                         \\
		                   &
		x_{i+1|k}=f(x_{i|k},u_{i|k}),  \quad \forall i \in \{0,\ldots,T-1\}, \\
		                   &
		g(x_{i|k},u_{i|k}) \leq 0, \quad \forall i \in \{0,\ldots,T-1\},     \\
		                   & x_{T|k}\in\mathcal S_f ,
	\end{aligned}
\]
where $T$ is the horizon and $\mathcal S_f$ denotes a terminal safe set for which safety can be guaranteed beyond the prediction horizon through a local controller. \edits{Here, $x_{i|k}$ and $u_{i|k}$ denote the state and input predicted $i$ steps ahead of time $k$, with $x_{0|k}=x_k$.}

The filter applies the first input $u_{0|k}^\star$ of the optimal sequence: whenever the desired input admits a feasible backup trajectory, it is passed through unmodified; otherwise, it is minimally adjusted. \edits{Recursive feasibility and closed-loop constraint satisfaction are guaranteed, provided that the terminal set $\mathcal S_f$ is (control) invariant.} \edits{In contrast to HJ reachability and CBFs, the safe set is not computed explicitly. The terminal set $\mathcal S_f$ can be small and simple, e.g., an invariant set of a local linear controller, and the region from which safety is certified is enlarged by the online planning over the horizon $T$. The computational effort is thereby shifted from the offline set computation to the online optimization.}

A closely related approach is shielding~\cite{alshiekh2018safe, bastani2021safe}, originally proposed in the context of \ac{RL}, in which a supervisory mechanism overrides the learned policy whenever it would lead to unsafe behaviour. In particular, model predictive shielding can be interpreted as a simulation-based variant of the \ac{PSF}: rather than solving an optimal control problem online, it forward-simulates a fixed backup policy and verifies that from the successor state there exists a feasible trajectory under the backup policy that ensures future constraint satisfaction. If this is not the case, the backup policy is applied instead.

\edits{The three approaches thus differ with respect to the representation of the safe set: it is computed explicitly by HJ reachability, characterized by a barrier function in the case of a CBF, and defined implicitly by the existence of a feasible backup trajectory in the case of a \ac{PSF}.}

\edits{The complexity of the online problem differs between the approaches. For HJ reachability, the filter only requires the evaluation of the precomputed value function~\cite{borquez2024safety}. The CBF filter has a single one-step constraint. It is convex if $b\circ f$ is concave in $u_k$, e.g., for linear dynamics and a concave barrier function; otherwise, it may be a small nonconvex program. In continuous time and for control-affine dynamics, the barrier constraint is affine in the input and the filter is a \ac{QP}~\cite{ames2019control}. The PSF is a finite-horizon optimal control problem with $T$ stages. It is a \ac{QP} for linear dynamics with polytopic constraints and terminal set, and a nonconvex nonlinear program for nonlinear dynamics.}

\subsection{\texorpdfstring{\edits{Robust and Stochastic Invariance}}{Robust and Stochastic Invariance}}
\edits{The notion of invariance extends to uncertain systems,} where it must hold for all admissible disturbances in robust settings~\cite{blanchini1999invariance}, and in probability or in expectation in stochastic ones~\cite{kofman2012probabilistic,hewing2018correspondence}.

In the robust case, \ac{HJ} reachability extends to worst-case analysis by computing invariant sets under adversarial disturbances~\cite{choi2023forward,seo2022real}, robust \acp{CBF} incorporate the uncertainty directly into the barrier conditions~\cite{xu2015robustness, choi2021robust}, and \acp{PSF} account for it through model-based constraint tightening mechanisms~\cite{wabersich2018linear, wabersich2021predictive}.
\edits{In the stochastic case, stochastic reachability methods characterize the probability of remaining within the safe set~\cite{abate2008probabilistic},} barrier-based approaches admit stochastic variants with prescribed confidence levels~\cite{cosner2023robust}, and \acp{PSF} extend through chance-constrained or risk-based optimization formulations~\cite{wabersich2021probabilistic}.

\subsection{\texorpdfstring{\edits{Relaxed Safety Formulations}}{Relaxed Safety Formulations}}
An alternative formulation of safety, particularly common in \ac{RL}, replaces hard constraint satisfaction by bounds on expected safety costs. \edits{This leads to \acp{CMDP}~\cite{altman2021constrained},} in which the control objective is optimized subject to constraints of the form
\[
	\mathbb E\!\left[\sum_{k=0}^{\infty} c(x_k,u_k)\right]
	\le d ,
\]
where $c$ measures safety violations and $d$ denotes an allowable constraint budget. Such formulations trade strict invariance guarantees for long-term probabilistic or average-performance guarantees.

In practice, strict constraint satisfaction may be infeasible or excessively conservative in the presence of uncertainty and disturbances. For this reason, many safety formulations employ soft constraints~\cite{nocedal2006numerical}, which permit temporary constraint violations while preserving optimization feasibility.

Consider the constraints~\eqref{eq:constraints}. A soft-constrained formulation introduces a nonnegative slack vector $\varepsilon_k \geq 0$ and relaxes the constraints to
\[
	g(x_k, u_k) \leq \varepsilon_k ,
\]
where each component of $\varepsilon_k$ specifies the permitted violation of the corresponding constraint. The resulting optimization problem  penalizes constraint violations through an augmented objective of the form
\[
	J_{\mathrm{soft}}
	=
	J
	+
	\lambda_1 \|\varepsilon\|_1
	+
	\lambda_2 \|\varepsilon\|_2^2 ,
\]
where $\varepsilon$ denotes  the slack vectors collected over the optimization horizon, and $\lambda_1, \lambda_2 \ge 0$ determine the trade-off between performance and constraint violation. Linear and quadratic penalties are typically combined: the linear term keeps the penalty exact, while the quadratic term penalizes large violations more strongly. In convex settings, choosing $\lambda_1$ sufficiently large makes the penalty exact, i.e., the soft-constrained problem recovers the solution of the original hard-constrained problem whenever the constraints can be satisfied, and violations occur only when the hard-constrained problem is infeasible. \\

\begin{tutorialexample*}[label={ex:safety-filter}, title={\tutorialtitlelabel~\thetcbcounter}]
	{Stability and safety filters}{}
	The relevant notion of stability depends on the assumptions on the system and its uncertainty. Nominal, robust and stochastic settings therefore lead to different stability guarantees, which are, however, only valid under the respective assumptions. In learning-based control, the characterization of the model error determines the setting, since the error enters the analysis either as a bounded disturbance or as a probabilistic uncertainty description.

	By means of a safety filter, an arbitrary control law, e.g., a learned policy, can be augmented with guarantees on constraint satisfaction. Hamilton-Jacobi (HJ) reachability, control barrier functions (CBFs) and predictive safety filters (PSFs) are all based on the notion of invariance, but differ with respect to conservativeness, scalability and computational effort. The choice among the three approaches is dependent on the particular problem: HJ reachability is attractive when the state dimension is small, CBFs when a valid barrier function can be constructed and PSFs when the computational effort can be shifted to the online optimization. Recent results focus on combining these techniques in order to leverage their individual benefits and to address their limitations using learning~\cite{wabersich2023data}. A detailed design example and comparison of the three different safety filter concepts for an inverted pendulum problem can be found in~\cite{wabersich2023data}. Note that filtering concepts can also incorporate stability requirements  by incorporating suitable Lyapunov decrease conditions, see e.g. ~\cite{DIDIER2024200,10886293}.

	Various software implementations are available for the presented safety filter concepts. In the following, a selection of commonly used tools is given, which is, however, not intended to be complete. For \ac{HJ} reachability, the value function can be computed on a grid of the state space using, e.g., the Level Set Toolbox~\cite{mitchell2007toolbox}, hj\_reachability~\cite{schmerling2021hjreachability} or OptimizedDP~\cite{bui2022optimizeddp}. DeepReach~\cite{bansal2021deepreach} approximates the value function by means of a neural network and is thereby applicable to systems of higher state dimension. Reachable sets based on set representations can be computed with, e.g., CORA~\cite{althoff2015cora} or JuliaReach~\cite{bogomolov2019juliareach}.

	The online evaluation of a \ac{CBF} filter amounts to the solution of a small optimization problem, which, if it can be formulated as a \ac{QP}, can be solved using standard solvers such as HPIPM~\cite{frison2020hpipm}, OSQP~\cite{stellato2020osqp}, Clarabel~\cite{goulart2024clarabel} or qpOASES~\cite{ferreau2014qpoases}. Note that the construction of a valid barrier function is in general not automated. In the following, some tools are given which support this step. For polynomial systems, barrier certificates can be obtained by means of sum-of-squares programming~\cite{prajna2004safety} using, e.g., SOSTOOLS~\cite{papachristodoulou2021sostools}. FOSSIL~\cite{abate2021fossil} synthesizes Lyapunov functions and barrier certificates represented by neural networks and formally verifies the resulting candidates, whereas refineCBF~\cite{tonkens2022refining} refines a given candidate function by means of \ac{HJ} reachability such that a valid \ac{CBF} is obtained. CBFkit~\cite{black2024cbfkit} provides implementations of \ac{CBF}-based safety filters for robotic applications and, in addition, allows a candidate barrier function to be fitted to labeled data by means of a neural network, the validity of which has to be verified separately.

	The online problem of a \ac{PSF} is a finite-horizon optimal control problem, which can be formulated using, e.g., CasADi~\cite{andersson2019casadi} and solved by a general-purpose nonlinear programming solver such as IPOPT~\cite{wachter2006ipopt}. Alternatively, frameworks that are specifically designed for real-time optimal control, see, e.g., acados~\cite{verschueren2022acados} and FORCESPRO~\cite{zanelli2020forcesnlp}, can be applied.

\end{tutorialexample*}
\section{Dynamics Learning for Control}\label{sec:dynamics}
Many control methods use a dynamics model to synthesize a controller for the problem formulated in~\autoref{sec:problem_formulation} and to analyze the resulting closed-loop properties, as described in~\autoref{sec:safety}. The achievable performance and guarantees depend on the accuracy of the model, and on how the associated uncertainty is accounted for. In practice, dynamics models may be constructed from our prior knowledge (e.g., first-principles physical models) for some applications; for others, such models may be unavailable or may not adequately capture complex dynamics. This has motivated a large body of work in learning-based control that incorporates measured data to learn or refine dynamics models, thereby complementing incomplete prior knowledge. Within this body of work, the learned models are leveraged to improve control performance, while control-theoretic analysis provides the foundation for retaining desired closed-loop properties.

To facilitate our discussion, we distinguish between \textit{direct} and \textit{indirect} data-driven control. In indirect data-driven control, a system model is distilled based on data collected from the real system, and this model is then used for controller design~\cite{sys_id}. Direct data-driven control approaches, on the other hand, map system observations to control actions based on past data, without explicitly defining a model.  In this section, we focus on the former and provide an introduction to common dynamics learning approaches, together with how they have been integrated with control frameworks. Methods for direct data-driven control are described in greater depth in~\autoref{sec:ddc}. In the following subsections, we first formulate the dynamics learning problem and then review common methodological threads, ranging from linear dynamics learning to deterministic nonlinear dynamics learning, and to probabilistic nonlinear dynamics learning approaches. 

\subsection{Dynamics Learning Problem}
We consider a discrete-time system of the form~\eqref{eq:system}, for which the dynamics can be partially known or fully unknown. Our goal is to infer a model of the dynamics from a sequence of measured inputs and states of length $N$, as defined in~\eqref{eq:data_u}, \eqref{eq:data_x}, and \eqref{eq:data_x+}. For the discussion below, we consider systems subject to additive process noise and approximate the true dynamics with a model $\hat{f}$ of the form
\begin{equation} \label{eq:learnablemodel}
\hat{f}(x,u) = \hat{f}_n(x, u) + \hat{f}_l(x, u),
\end{equation}
where $\hat{f}_n$ is a nominal model that captures available prior knowledge, and $\hat{f}_l$ is the component to be learned from data. When the system dynamics are fully unknown, the nominal model $\hat{f}_n$ can be set to the zero function. For convenience in the following discussion, we define an augmented vector $z=[x^\top,u^\top]^\top$, which is the input to both model components.

Over the years, many works have considered dynamics learning problems of the form~\eqref{eq:learnablemodel}. Beyond incorporating a prior dynamics model, dynamics learning can also leverage structural knowledge of the system~\cite{umlauft2018uncertaintybased,evangelisti2022physically,greeff2021exploiting,carron2019data, frohlich2022contextual} to improve learning efficiency or model accuracy. We categorize the resulting frameworks according to the formulation of the learned model, distinguishing between linear and nonlinear models as well as deterministic and probabilistic formulations. The chosen formulation influences how model uncertainty can be characterized and accounted for during closed-loop operation. In the following, we provide an overview of representative methods, with particular emphasis on the learning framework, the controller synthesis methodology, and the guarantees that can be provided under possible dynamics uncertainties. \edits{A summary of the different learning methods, along with their advantages, disadvantages, compatible control methods, and the stability/safety guarantees they enable, is given in \autoref{tab:method-comparison}. Note that the control methods provided in the table are not exhaustive, but are only intended to serve as a starting point for those new to the field. }

\subsection{Learning Linear System Models}
\label{subsec:linear}
We begin our discussion with linear dynamics learning problems, where the true dynamics can be written as
\begin{equation} \label{eq:linearmodel}
x_{k+1} = Ax_k + Bu_k+w_k
\end{equation}
with $A$ and $B$ being system matrices of appropriate dimensions. For simplicity, we first suppose that the nominal component is absent, and the approximated dynamics model analogously takes a linear form: $x_{k+1} = \hat{A}x_k + \hat{B}u_k$. While our focus here is on linear dynamics, for nonlinear systems, linear models can also be used to approximate the local dynamical behaviour around an operating point. Such linear representations can simplify controller synthesis, which can be desirable in certain applications where the approximation error is sufficiently small.

The model parameters $\Theta = [\hat{A}, \hat{B}]$ are typically determined using least-squares regression (or its variants). Specifically, the learning problem may be formulated as the following optimization problem: $\arg\min_{\Theta} \| X^+ - \Theta M \|_F^2$, where $M = [X^\top,\: U^\top]^\top$. If $M M^\top$ is invertible, we can find the closed-form solution $\Theta = X^+M^\top(M M^\top)^{-1} $. This formulation can be related to the maximum likelihood estimation problem (i.e., maximizing the probability of observing the data given the model parameters) under appropriate noise assumptions~\cite{least_squares_mle}. To generalize to the case when the nominal model~$\hat{f}_n$ is available, the closed-form solution can be obtained by adapting the datasets to capture the residual dynamics $x_{k+1} - \hat{f}_n(x_k, u_k)$. 

\edits{Note that the computational complexity of linear regression is linear in the number of data samples. However,} even for systems with true linear dynamics, convergence to true parameters can require a large amount of data; it was shown in~\cite{tsiamis2021linear} that, for certain cases, the worst-case sample complexity is exponential. Nevertheless, finite-sample error bounds can be estimated~\cite{dean_sample_2020,abbasi2011regret}. These can be norm-bounded estimation errors (e.g., $\|\Delta_A\| \leq e_A$ and $\|\Delta_B\|\leq e_B$ with $\Delta_A =  A -  \hat A$ and $\Delta_B =  B - \hat B$ ), or in the form of confidence ellipsoidal sets where
\begin{equation*}
\text{trace}\left (\begin{bmatrix} 
\Delta_A & 
\Delta_B
\end{bmatrix}Q_p  \begin{bmatrix}
\Delta_A^\top \\ 
\Delta_B^\top
\end{bmatrix} 
\right ) \leq c
\end{equation*}
holds with probability at least $(1-\delta)$ for some positive definite matrix $Q_p$ and positive scalar \edits{$c$ that depends on a user-chosen confidence parameter $\delta$}. 

The primary objective of the uncertainty bounds obtained above is to inform the closed-loop controller synthesis, which may be performed for linear systems using well-established techniques such as \ac{LQR} and \ac{MPC}. When using LQR, it is shown that a stabilizing controller can be determined with a suboptimal guarantee with estimated dynamics~\cite{mania_certainty_2019}. Alternatively, if the errors are large, then one can also use robust LQR tools (e.g., system-level synthesis~\cite{anderson_system_2019}). 
For MPC-based controller synthesis in the presence of norm-bounded error estimates, one can rewrite the system dynamics as $x_{k+1} = \hat Ax_k + \hat Bu_k+\tilde w_k$, where $ \tilde w_k= (\Delta_A)x_k + (\Delta_B)u_k + w_k$. If $x_k\in\mathcal X \subseteq \mathbb R^{n_\mathrm{x}}$, $u_k\in\mathcal U \subseteq \mathbb{R}^{n_\mathrm{u}}$, and $w_k\in\mathcal W \subseteq \mathbb{R}^{n_\mathrm{w}}$, an aggregate disturbance set $\widetilde{\mathcal W}$ can be constructed such that $\Delta_A\mathcal X\oplus\Delta_B\mathcal U\oplus\mathcal W \subseteq \widetilde{\mathcal W}$. 
The resulting set can then be accounted for using constraint tightening or tube-based robust MPC methods~\cite{mayne_robust_2005}.
Similarly, when confidence bounds for the learned dynamics are available, robust MPC methods can be extended to ensure constraint satisfaction in the probabilistic sense or can be combined with stochastic \ac{MPC}. 
This will be further discussed when dealing with probabilistic models in~\autoref{subsec:probabilistic}.

Safety filters may also be employed (e.g.,~\cite{wabersich2018linear}) to ensure robust constraint satisfaction during closed-loop operation. 
For unknown linear systems subject to \edits{(sub-)}Gaussian noise, learning-based safety filters use estimates of the dynamics and noise covariance to ensure constraint satisfaction with high probability~\cite{farokhi_learning_2023, self_tuning}.
Related efforts investigate when multi-step prediction can reduce compounding model errors~\cite{somalwar2025learning} and develop finite-sample guarantees for learning and control~\cite{finite_sample}.

\subsection{Learning Deterministic Nonlinear Models} \label{subsec:nn}
Having introduced linear dynamics learning, we now turn to nonlinear systems. While local linear approximations can be effective in some applications, their accuracy may be limited for systems with strongly nonlinear dynamics. Learning nonlinear dynamics models from data provides a means to capture more complex system behaviour for further improving closed-loop performance. Learning general nonlinear models of the form~\eqref{eq:learnablemodel} without any prior structural parameterization can be inefficient. We can often leverage a priori knowledge of the system (e.g., leveraging differential flatness,  control-affine structures, or polynomial dynamics) to  facilitate the learning process.

One common set of approaches in the literature involves learning residual dynamics with  linear regression:
\begin{equation} \label{eq:linparammodel}
x_{k+1} = \hat{f}_n(z_k) + \Theta \phi(z_k),
\end{equation}
where $\phi(z_k)$ is a vector of basis functions capturing system nonlinearities, and $\Theta$ is the parameter matrix to be learned. Note that, while the model in~\eqref {eq:linparammodel} is nonlinear in $z_k$, it is linear in the parameter matrix $\Theta$; consequently, under the corresponding rank condition, the least-squares estimate of~$\Theta$ admits a closed-form solution, similar to the case in~\autoref{subsec:linear}.

When prior knowledge of the system is limited, nonlinear dynamics can instead be represented using neural networks, which are often adopted because of their universal approximation capabilities~\cite{hornik_multilayer_1989}. In a typical setting (e.g.,~\cite{bansal2016learning,zhou2017design}), the learnable model $\hat f_{l, \theta}$ is a feedforward network and is defined as a sequence of affine and nonlinear transformations~\cite{bishop2006pattern}:
\begin{equation} \label{eq:nn}
\begin{cases}
    z^0 = z , \\
    z^{i+1} = \mathrm{act}(W^iz^i + b^i), \quad i \in \{0,\ldots,l-1\} \\
   z^{l+1} = W^lz^{l} + b^l.
\end{cases}
\end{equation} 
Here, $l$ denotes the number of hidden layers in the neural network, $z^0=z$ is the input of the network, $z^{l+1}$ is the output of the network, $\theta$ denotes the augmented vector of the learnable parameters $W^i$ and $b^i$ across all layers, and $\mathrm{act}(\cdot)$ is a nonlinear activation function (e.g., ReLU or tanh). The goal of the training process is to reduce the error between the predicted output and the target in the dataset. One commonly used objective is the mean-squared error:
\begin{equation}
\min_{\theta} \:\frac{1}{N} \sum_{j=0}^{N-1} \left \| x_{j+1} - \big(\hat{f}_n(z_j) + \hat f_{l,\theta}(z_j)\big )\right \|^2,
\end{equation}
where the subscript $j$ denotes the $j$-th data point with $z_j=[x_j^\top,u_j^\top]^\top$ and $x_{j+1}$ being the corresponding successor state.
Beyond standard feedforward neural networks, nonlinear dynamics learning may also employ frameworks such as neural ODEs~\cite{chen_neural_2018}, physics-informed neural networks~\cite{nghiem2023physics, raissi_physics-informed_2019}, and neural autoregressive models~\cite{connor_recurrent_1991} to incorporate richer dynamical structure, physical knowledge, or temporal dependencies. Learned neural network dynamics models have been combined with nonlinear MPC~\cite{armenio2019model, bonassi2022offsetfree, giuli2024modeling, ravasio2024lmibased} and safety filters~\cite{wang2024providing}. 

While neural networks are expressive, their black-box nature makes it more difficult to provide closed-loop guarantees. Nevertheless, recent advances have attempted to address this challenge by exploiting the inherent properties of learned models. One property commonly leveraged in closed-loop analysis is Lipschitz continuity. For feedforward neural networks with a finite number of layers and globally Lipschitz-continuous activation functions, there exists a constant $L\ge 0$ such that $\| \hat f_{l,\theta}(z) - \hat f_{l,\theta} (z') \| \leq L\|z - z' \|$, $\forall z, z'\in \mathbb{R}^{n_\mathrm{x}+n_\mathrm{u}}$. Intuitively, the Lipschitz property allows us to bound the change in the network output induced by variations or perturbations in its input. While estimating the Lipschitz constants can be non-trivial~\cite{lipschitz_nn}, Lipschitz constraints can be encouraged or enforced through training procedures~\cite{miyato2018spectral,pauli_training_2022,bungert_clip_2021} or directly through the network architecture design~\cite{anil2019sorting}. A representative training-based approach is spectral normalization~\cite{miyato2018spectral}, in which the weight matrix of each layer is rescaled during training to constrain its spectral norm. If the activation functions are 1-Lipschitz with respect to the Euclidean norm, as in the case of many common activation functions (e.g., ReLU and tanh), then an upper bound on the network Lipschitz constant is given by $\prod_{i=0}^{l}\|W^i\|_2$. Beyond spectral normalization and related weight-projection methods, Lipschitz continuity can also be imposed through network architectures that combine norm-constrained linear mappings with gradient-norm-preserving activation functions~\cite{anil2019sorting}. Both spectrally normalized networks and Lipschitz-constrained architectures have been integrated with control-theoretic tools, including Lyapunov-based nonlinear control and small-gain arguments, to establish closed-loop stability guarantees for learning-enhanced systems in practical applications~\cite{shi2019neural,zhou_bridging_2022}. Besides Lipschitz continuity, neural networks can also be characterized through local sector bounds or quadratic constraints on their activation functions; these descriptions preserve more information about the network structure than a single global Lipschitz bound and can be incorporated into Lyapunov- and robust-control analyses (e.g.,~\cite{pauli2021linear,yin_stability_2022}) to establish desired closed-loop properties.

While the preceding approaches are well-suited to analyzing feedforward neural networks as static input-output mappings, many learning approaches utilize autoregressive models, recurrent neural networks, and neural ODEs, which naturally define dynamical models with internal states or temporal dependencies induced by state-input sequences of the form~\eqref{eq:data_u}, \eqref{eq:data_x}, and \eqref{eq:data_x+}. For these models, it is important to analyze how their internal states and output trajectories evolve over time. This can be done by studying properties such as ISS and $\delta$-ISS and using them in the closed-loop analysis to establish convergence of the system trajectories to the desired behaviour or to a bounded neighbourhood thereof in the presence of model mismatch (see~\autoref{sec:safety}). In this regard, sufficient conditions for ISS and $\delta$-ISS properties have been obtained for neural autoregressive models~\cite{bonassi_stability_2021} and recurrent neural networks~\cite{bonassi_lstm_2020, amico_iss_rnn_2024}. Related contractivity and dissipativity guarantees have been developed for neural ODEs~\cite{martinelli2023unconstrained}, while stable-by-design recurrent networks are proposed in~\cite{revay_recurrent_2024}. These model properties can be incorporated directly into controller analysis and synthesis. For instance, in~\cite{armenio2019model}, $\delta$-ISS properties are leveraged for the derivation of MPC laws ensuring stability and recursive feasibility. This has also been extended to offset-free tracking control in~\cite{bonassi2022offsetfree}.

\subsection{Learning Probabilistic Nonlinear Models}
\label{subsec:probabilistic}
Although nonlinear dynamics learning methods are more flexible than their linear counterparts, this flexibility can come at the cost of conservatism due to the assumed worst-case bounds when handling model uncertainty. While methods such as approximate Bayesian inference~\cite{bao_epistemic_2021, gal2016dropout}, approximate Cram\'er–Rao lower bound~\cite{cohn1993neural}, and deep ensembles~\cite{lakshminarayanan_simple_2017} exist, accurately and reliably estimating predictive uncertainty for highly expressive function approximators, such as neural networks, remains a non-trivial problem.  For closed-loop control, one may alternatively leverage probabilistic learning methods, which allow us to systematically bookkeep model uncertainties. Such probabilistic characterizations enable the use of chance constraints or risk-sensitive formulations in place of purely worst-case constraints, providing a principled means to trade off control performance and confidence in constraint satisfaction.

To make the problem tractable, i.i.d.\ Gaussian assumptions are often imposed on the process or measurement noise. For a model that is linear in the parameters, e.g., of the form~\eqref{eq:linparammodel}, and with a Gaussian prior $\theta \sim \mathcal{N}(\mu_\theta, \Sigma_\theta)$ on the unknown parameter vector $\theta \in \mathbb{R}^{n_\phi}$  for a single output dimension, the posterior distribution over $\theta$ is again Gaussian and available in closed form~\cite{bishop2006pattern}; this is known as \ac{BLR}. In the single-output case, the posterior mean and covariance are $\hat{\mu}_\theta = \hat{\Sigma}_\theta \left(\Sigma_\theta^{-1}\mu_\theta + \sigma_w^{-2}  M_\phi X_+ ^\top \right)$ and 
    $\hat{\Sigma}_\theta^{-1} = \Sigma_\theta^{-1} + \sigma_w^{-2} M_\phi M_\phi^\top$,
where $M_\phi = [\phi(z_0)\ \ldots\ \phi(z_{N-1})] \in \mathbb{R}^{n_\phi \times N}$. The multi-output case follows by applying these expressions independently to each row of $\Theta$.

Apart from \ac{BLR}, nonlinear dynamics can also be represented using non-parametric probabilistic models such as \acp{GP}~\cite{williams2006gaussian}.  
When approximating $\hat f_l$ with a \ac{GP}, we assume that its function values are jointly Gaussian, with covariance specified by a kernel function $\kappa$. For illustration, we  consider a single output  and define the residual training targets as $\Gamma^\top = X^+-\hat F_n(M)$, where $\hat F_n(M)$ denotes the nominal-model predictions evaluated at the training inputs. Then, for a new test point~$z=[x^\top,u^\top]^\top$, we have
\begin{equation} \label{eq:gpr}
\begin{bmatrix}
\Gamma \\
\hat f_l(z)
\end{bmatrix} \sim \mathcal{N}\bigg(0, \begin{bmatrix} K+ \Sigma^w & K_{z} \\ K^\top_{z} & \kappa(z, z) \end{bmatrix} \bigg),
\end{equation}
where $K$ is the Gram matrix computed from data $(X, U)$ using $\kappa$, $K_z$ contains the kernel evaluations between the training inputs and the test input $z$, and $\Sigma^w$ denotes the covariance matrix of the noise term. Conditioning on the observed residual targets yields $p\left(\hat f_l(z)\mid Z,\Gamma\right) = \mathcal{N}(\hat \mu, \hat \Sigma)$, where $\hat\mu=K_z^\top\left(K+\Sigma^w\right)^{-1}\Gamma$ and $\hat\Sigma=\kappa(z,z)-K_z^\top\left(K+\Sigma^w\right)^{-1}K_z.$ The posterior mean of the full dynamics can then be expressed as $\hat f_n(z)+\hat\mu(z)$.

\edits{In practice, BLR and GPs offer complementary advantages and limitations.  For both model classes, reliable predictive uncertainty estimates depend on appropriate prior structure, encoded through the choice of basis functions in \ac{BLR} and the kernel and its hyperparameters in \acp{GP}. This dependence is illustrated in \autoref{ex:model-learning}. Moreover, computationally, fitting a \ac{GP} typically scales cubically with respect to the number of data samples, while \ac{BLR} scales linearly, offering an advantage over \ac{GP}~\cite{computation_gpblr}, at the cost of representational flexibility. While this computational tradeoff may not be critical in offline learning settings, online learning settings often require a careful balance between fast adaptation and model capacity.
}

\begin{tutorialexample*}[label={ex:model-learning}]
  {Learning the unknown}
  {http://tiny.cc/cdc-tut-ex2}
When the system dynamics are not fully known, learning can help identify unknown components and thereby improve control performance. Reliable learning depends on both informative data and appropriate prior structure. Observed data constrain the model within sampled regions, while prior structure shapes its generalization beyond them. With two  probabilistic learning approaches, this example illustrates how data and structure jointly influence predictive accuracy and the reliability of the associated uncertainty estimates. The  implementation for this example is available in \cite{cfr-notebook} (\href{http://tiny.cc/cdc-tut-ex2}{BLR vs. GP}).

We consider a car traversing unknown terrain, represented by an elevation profile~$h(p)$, where $p$ denotes the car’s horizontal position and $h(p)$ denotes the corresponding terrain height. For this example, the true terrain profile is $h(p)=0.01\cos(18 p)$  and is observed through noisy measurements $\mathcal{D}= \{(p^{(d)},h^{(d)})\}_{d=1}^{D}$, where the superscript $^{(d)}$ denotes the data index, $h^{(d)} = h(p^{(d)})+\varepsilon^{(d)}$, and $\varepsilon^{(d)}\sim\mathcal{N}(\mu_\varepsilon,\sigma_\varepsilon^2)$ is \ac{iid} Gaussian noise with  mean $\mu_\varepsilon=0$ and variance $\sigma_\varepsilon^2=2.5\times 10^{-5}$. Our goal is to learn the function $h(p)$ using the dataset $\mathcal{D}$.

\autoref{fig:tutorial-model-learning} compares the learning performance of \ac{BLR} and \ac{GP}, when different structural assumptions were made. From the top-left panel, we see that, with exactly matching periodic basis functions, the predictive mean and variance of the \ac{BLR} (red) generalize reliably, even across the unsampled data gap (grey). As further illustrated in the bottom-left panel, when such task-specific structure is unavailable, a \ac{GP}, with its hyperparameters tuned via maximizing the log marginal likelihood, can be a flexible alternative, though \tutorialcolumnbreak  with increased uncertainty bounds over the unsampled data region.  
Overall, both methods rely on structural assumptions (e.g., basis functions for BLR and the kernel and its hyperparameters for GPs). The right-hand panels illustrate the consequences of model misspecification. An incorrect basis or  kernel hyperparameters can bias the predictive mean and produce  overconfident predictions. Reliable generalization depends on appropriate structural assumptions, especially when only a limited number of observations is available.

\begin{tutorialfigure}
    \includegraphics[width=\linewidth]
        {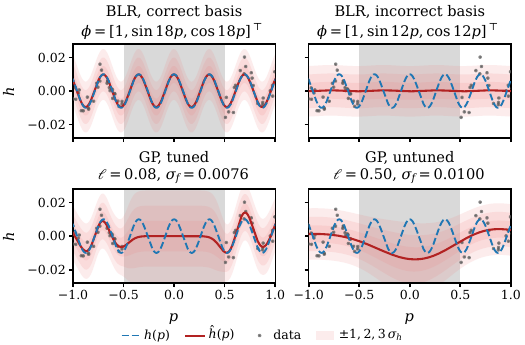}
    \captionof{figure}{
        Effect of prior structure on learning from incomplete data. The grey region contains no measurements; the dashed blue and solid red curves denote the true profile and posterior mean, respectively, and the shaded bands correspond to model predictive uncertainties.
    }
    \label{fig:tutorial-model-learning}
\end{tutorialfigure}
\end{tutorialexample*}

It should be noted that the \edits{learned probabilistic models} only make one-step predictions rather than trajectory predictions. This can be challenging in closed-loop control methods such as MPC, where dynamics uncertainty must be propagated over the prediction horizon. Common approaches approximate the distributions of subsequent predicted states using methods such as linearization~\cite{hewing_simulation_2020}, moment matching~\cite{opper_bayesian_1999}, and sigma-point propagation~\cite{sigma_point}. Depending on the formulation, additional simplifying assumptions may be introduced, such as neglecting correlations between successive GP evaluations. Starting from a state distribution at time $k$ with mean $x_k^\mu$ and covariance $\Sigma_k^x$, these methods yield approximate predicted state distributions characterized by $(x_{i\mid k}^\mu,\Sigma_{i\mid k}^x)$ for future prediction steps $i>k$.

Closed-loop guarantees can then be pursued by solving a stochastic optimal control problem with chance constraints, \edits{both in the case of \ac{BLR} and \ac{GP}~\cite{computation_gpblr, scampicchio_gaussian_2025}.} For example, a joint chance constraint over a finite prediction horizon may be written as $p\big(g(x_{i\mid k},u_{i\mid k})\leq0,\ \forall i\big) \geq \alpha$, where $\alpha$ is a user-defined confidence level. To obtain a tractable controller synthesis problem, such probabilistic constraints are commonly replaced by deterministic reformulations or conservative approximations of the form
$g(x_{i\mid k}^\mu,u_{i\mid k})+\nu(x_{i\mid k}^\mu,u_{i\mid k},\Sigma_{i\mid k}^x)\leq 0,$ where $\nu(\cdot)$ represents a constraint tightening determined by the predicted state uncertainty, the assumed probability distribution, and the selected confidence level.
The resulting optimization problem can then be solved (e.g., via MPC) as
\begin{subequations}
\label{eqn:pl_opt_ctrl}
\begin{align}
    \min_{u_{i\mid k}} \quad & J \\
    \text{s.t. } \quad & x^\mu_{0\mid k} = x_k, \\
    & \Sigma^x_{0 \mid k} = 0,\\
    &x^\mu_{i+1\mid k} = \rho(x^\mu_{i \mid k}, \Sigma^x_{i \mid k}, u_{i \mid k}), \\
    & \Sigma^x_{i+1\mid k} = \Lambda(x^\mu_{i \mid k}, \Sigma^x_{i \mid k},u_{i \mid k}), \\
    & g(x^\mu_{i \mid k},u_{i \mid k}) + \nu(x^\mu_{i \mid k}, u_{i \mid k}, \Sigma^x_{i \mid k}) \leq 0, \\
    & \forall i \in \{0,\ldots, T-1\}, \nonumber
\end{align}
\end{subequations}
where $x_k$ is the current state, $\rho(\cdot)$ and $\Lambda(\cdot)$ are the propagation of the predicted state mean and covariance  under the chosen uncertainty-propagation method. Note that the covariance propagation significantly increases the computational complexity of the optimization problem. The computational effort can be reduced by employing SQP methods with tailored Jacobian approximations~\cite{lahr2023zero}, for which an efficient open-source implementation is provided in~\cite{lahr2026l4acados}. \edits{Moreover, an ancillary controller can be introduced to  reduce deviations from the predicted nominal trajectory, thereby reducing uncertainty growth over the prediction horizon and the associated constraint tightening.} A review of MPC with GP models can be found in~\cite{scampicchio_gaussian_2025}, \edits{and \autoref{ex:gp-mpc} shows an example of GP-MPC.}\\

\begin{tutorialexample*}[label={ex:gp-mpc}]
    {Acting under uncertainty}  {http://tiny.cc/cdc-tut-ex3}
    The previous example focused on learning the unknown terrain profile and quantifying its predictive uncertainty. We now close the loop by embedding the learned \ac{GP} model within an \ac{MPC} controller. This example illustrates how the resulting dynamics uncertainty can inform control decisions and constraint tightening, enabling a systematic trade-off between closed-loop performance and confidence in constraint satisfaction. The  implementation of this example can be found in \cite{cfr-notebook} (\href{http://tiny.cc/cdc-tut-ex3}{GP-MPC}). 

Building on the terrain model learned in \autoref{ex:model-learning}, we introduce the vehicle dynamics and control task; an illustration is shown in \autoref{fig:tutorial-gp-mpc}. The vehicle has state $x=[p,v]^\top\in\mathbb{R}^2$, comprising its horizontal position $p$ and velocity $v$, and a scalar control input $u\in [-8,8]$. The vehicle  dynamics are given by
  \begin{align}
  \dot p = v, \:\: \dot v = u\:\cos\big(\alpha(p)\big) - g\:\sin\big(\alpha(p)\big)\cos\big(\alpha(p)\big),
  \end{align}
  where $g$ denotes the gravitational acceleration and
$\alpha(p)=\arctan\bigl(dh(p)/dp\bigr)$ is the local terrain inclination angle. The task is to reach a target $p_\text{goal}$ subject to a speed limit $|v|\le v_{\max}=0.6$. The objective function is defined as $J=\sum_k 5(p_k-p_\text{goal})^2+0.01u_k^2$. We formulate the control problem in the form of~\eqref{eqn:pl_opt_ctrl} and solve it in a receding horizon manner. The nominal predictions $\rho(\cdot)$ are obtained based on  the \ac{GP} posterior mean prediction, while the state covariance $\Lambda(\cdot)$ is propagated through a linearized covariance recursion under an ancillary feedback controller. The predicted velocity uncertainty is incorporated by tightening the constraint as $  |v_{i|k}|+\beta\sigma^v_{i|k}\leq v_{\max}$, with $\beta\geq0$ controlling the degree of constraint tightening. In general, a larger $\beta$ provides a larger margin against model uncertainty, leading to greater confidence in constraint satisfaction. 

  The bottom-left panel of~\autoref{fig:tutorial-gp-mpc} shows the predicted state uncertainty and the corresponding constraint tightening at $t=0$ for $\beta=2$. The tightened constraint keeps the predicted mean velocity below $v_{\max}$ by a margin proportional to its predicted standard deviation. The grey ellipses show that, without ancillary feedback, the uncertainty grows more rapidly and would therefore require more restrictive constraint tightening. 
 The bottom-right panel shows the trade-off between \tutorialcolumnbreak  closed-loop cost and empirical constraint violations as $\beta$ is varied. Without constraint tightening ($\beta=0$),  
 the nominal controller exceeds the speed limit. Increasing~$\beta$ to 2 
 mitigates the observed violations  at the expense of a higher closed-loop cost. The green points further show that adding training data within the previously unsampled region reduces model uncertainty, allowing the learning-based \ac{MPC} controller to achieve a lower closed-loop cost at $\beta=2$, without  constraint violations.
 
  \begin{tutorialfigure}
      \includegraphics[width=\linewidth]
          {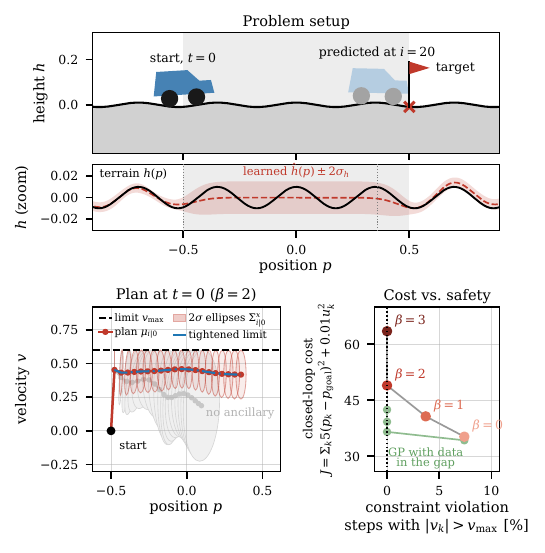}
      \captionof{figure}{Acting under model uncertainty with learning-based \ac{MPC}. The top panel shows the  setup for the car terrain traversal  example. The bottom-left panel shows the prediction with constraint tightening at $t=0$ for $\beta=2$, while the bottom-right panel shows the trade-off between closed-loop cost and constraint violation for different~$\beta$ choices.
      }
      \label{fig:tutorial-gp-mpc}
  \end{tutorialfigure}
  \end{tutorialexample*}

An alternative to the probabilistic uncertainty-propagation methods above is to use robust set-based propagation to construct ellipsoidal confidence regions for the predicted states~\cite{koller2018learningbased}. In this setting, the state $x$ is assumed to belong to an ellipsoidal uncertainty set $\mathcal{E}(x_p, Q_p)$ centred at $x_p$, with shape determined by $Q_p$. The system dynamics robustly propagate from one ellipsoidal set to another, and the sum of these ellipsoids gives the multi-step predictions of the system, i.e., they approximate the reachable sets. Then, e.g.,  a tube-based MPC can be formulated with set-based system dynamics for robust satisfaction of constraints. 

While these robust tools for closed-loop control guarantees provide worst-case constraint satisfaction, the confidence regions are often difficult to obtain, or, when available, too conservative in practice. A solution to this is to use sampling-based approaches for constraint tightening~\cite{prajapat2024towards, sampling_tighten} or using scenario-based MPC~\cite{scenario-mpc}. 
Separately, problems with GPs can also be solved using LQR via local linearizations~\cite{gp_robust, pmlr-v144-rohr21a}.

An important concern for closed-loop control with GPs is stability. Within the context of MPC, this can be guaranteed by imposing additional assumptions, for example, the existence of robust positively invariant sets~\cite{koller2018learningbased}, and requires careful selection of terminal constraints. Constructing them is often difficult in practice, so they are handled via sampling-based approaches, back-up controllers or soft constraints. Stability can also be addressed beyond the receding-horizon setting via Lyapunov functions~\cite{umlauft2018uncertaintybased}. Here, GP dynamics are slightly modified to ensure a known equilibrium, i.e., $f(0) = 0$, and fidelity on model knowledge is determined through uncertainty bounds. These uncertainty bounds are then incorporated to determine a \ac{CLF} and a stabilizing feedback law through dynamic programming. Lyapunov-based constructions have also been used for tracking control~\cite{beckers2017stable} and backstepping control design~\cite{capone_backstepping_2019}.

\subsection{Data Collection for Offline Versus Online Learning}
The dynamics learning methods discussed in this section are performed predominantly in an offline manner, i.e., a suitable model is learned based on a fixed, input-output dataset, and the control is designed based on the learned model prior to closed-loop deployment. The performance of such methods is tied to the assumption that the collected data is informative enough for estimation. In the machine learning literature, similar challenges have been addressed through active learning methods, which enable systematic data collection by actively querying the most informative data points~\cite{settles2012active}, and such techniques have been successfully integrated into learning-based control frameworks to improve sample efficiency and model accuracy~\cite{bansal2017goaldriven,zhou2019active, lee2024active}. In this context, common information criteria include \textit{(i)} mutual information, which quantifies the reduction in uncertainty about the unknown dynamics from the observations~\cite{capone_localized_2020, sukhija_optimistic_2023}, \textit{(ii)} entropy, which targets the regions in the state space where the system dynamics are the most uncertain~\cite{buisson-fenet_actively_2020} and \textit{(iii)} criteria based on the Fisher information matrix~\cite{wagenmaker_active_2020}, including D-optimality, which minimizes the volume of the parameter-confidence ellipsoid, and A-optimality, which minimizes the average parameter variance.

In contrast to offline learning, online learning approaches pair closed-loop control with online model adaptation through information acquisition. Although this is not the focus of our discussion, we note that active-learning methods also extend to online settings. Designing controllers that simultaneously achieve closed-loop performance and generate informative data for reducing model uncertainty is a central challenge in dual control~\cite{feldbaum_dualcontrol, arcari2020dual, arcari2020approximate}. Intuitively, the objective can often be decomposed as $J = J_p + J_i$, where $J_p$ is the performance objective (e.g., reference tracking costs) and $J_i$ is an information objective. Minimizing $J_i$ encourages probing control inputs that steer the system along trajectories expected to reduce uncertainty in the learned dynamics. Through a suitable choice of information objective, the controller can balance task performance with informative data collection and thereby improve the learning efficiency.

Many online-learning methods perform model adaptation at each control instant or prediction horizon. However, learning updates can be computationally costly, and adaptation at every step may be unaffordable or unnecessary. Approaches to mitigating this challenge include the use of a data-dictionary in combination with sparse GP methods~\cite{hewing2019cautious}, the exploitation of Kalman filters for online parameter updates~\cite{frohlich2022contextual}, spatio-temporal \acp{GP}~\cite{bartels2026real},  and \ac{ETL}~\cite{solowjow_event-triggered_2020,umlauft_feedback_2020,schluter_event-triggered_2021,zheng_nonparameteric_2023}, in which the learning task is performed when a triggering condition indicates that the current model is no longer sufficiently accurate.
\section{Direct Data-Driven Control} 
\label{sec:ddc}
An alternative approach to first learning the model of the system and then using it in a model-based control scheme is to directly use the data to design the desired controller. This approach is known as \textit{direct} data-driven control, as it avoids the intermediate step of system identification. Although diverse methods for direct data-driven control have been studied in the literature, most of the recent success in this area is based on---or at least inspired by---the result that is now known as Willems' fundamental lemma \cite{willems_lemma}. In this section, we focus exclusively on describing this approach.

In this section, we use the following notation. Given a sequence $u=\{u_k\}_{k=0}^{N-1}$, we define $u_{[a,b]}:=[u_a^\top~\dots~u_b^\top]^\top$ for $0\leq a \leq b\leq N-1$. Moreover, the Hankel matrix of depth $L$ of the sequence is
\begin{equation*}
		H_L(u) := \begin{bmatrix}
			u_0 & u_1 & \cdots & u_{N-L} \\
			u_1 & u_2 & \cdots & u_{N-L+1} \\
			\vdots & \vdots & \ddots & \vdots \\
			u_{L-1} & u_{L} & \cdots & u_{N-1}
		\end{bmatrix}.
	\end{equation*}

\subsection{Willems' Fundamental Lemma}
    
Willems' fundamental lemma describes how every input-output trajectory of a controllable discrete-time \ac{LTI} system can be characterized in terms of a single persistently excited trajectory measured from such a system. This result was first described using the \textit{behavioural approach} to systems theory in \cite{willems_lemma}, and has been replicated in the state space framework in \cite{DePersisTes2020,vanWaardePerCamTes2020,BerberichAll2020}. Willems' lemma has provided valuable new insights regarding the information contained in measured data, as well as in how to exploit it for system analysis and control design. In particular, a discrete-time LTI system of the form
\begin{subequations}
	\begin{align}
		x_{k+1} &= A x_k + B u_k, \label{dtsysx}\\
		y_k & = C x_k + Du_k,
	\end{align}
	\label{dtsyst}%
\end{subequations}
are considered\footnote{Nonlinear extensions to the fundamental lemma, as well as nonlinear data-driven controllers, have been obtained. We comment on them below.}. Then, a definition of persistence of excitation was proposed in \cite{willems_lemma} as follows. 

\begin{defn}[Persistently exciting sequence]
	\label{defpe}
	Consider a discrete-time input sequence $u = \{ u_k \}_{k=0}^{N-1}$, $u_k \in \mathbb{R}^{n_\mathrm{u}}$. The sequence $u$ is \ac{PE} of order $L$ if its Hankel matrix of depth $L$ has full row rank, i.e., $\text{rank} ( H_L(u) ) = n_\mathrm{u} L$.
\end{defn}

Applying a PE input to a linear system \eqref{dtsyst} allows one to collect input-output data that can be used to describe every other trajectory admitted by the system. In other words, a data-based system representation is obtained, as presented in the following theorem.

\begin{thm}[Willems' fundamental lemma \cite{willems_lemma}]
	\label{thmdtwill}
	Let system \eqref{dtsyst} be controllable and apply an input sequence $u=\{ u_k^d \}_{k=0}^{N-1}$ that is persistently exciting of order $L+n_\mathrm{x}$. Let $y=\{ y_k^d \}_{k=0}^{N-1}$ be the measured output trajectory of \eqref{dtsyst}. A pair of sequences $\{ \bar u_k \}_{k=0}^{L-1}$, $\{ \bar y_k \}_{k=0}^{L-1}$ is an input-output trajectory of \eqref{dtsyst} if and only if the expression
	\begin{equation}
		\begin{bmatrix}
			H_L(u^d) \\ H_L(y^d)
		\end{bmatrix} \alpha = \begin{bmatrix}
		\bar u \\ \bar y
		\end{bmatrix} 
		\label{dtwillems}
	\end{equation}
	holds for some vector $\alpha \in \mathbb{R}^{N-L+1}$, with $\bar u = [\bar u_0^\top \cdots \bar u_{L-1}^\top]^\top$ and $\bar y = [\bar y_0^\top \cdots \bar y_{L-1}^\top]^\top$.
\end{thm}

Now, \eqref{dtwillems} is a data-based representation of system \eqref{dtsyst} in the sense that any trajectory $\bar u, \bar y$ of length $L$ that can be generated by \eqref{dtsyst} can also be expressed in terms of the measured data $u^d,y^d$, by finding a suitable vector $\alpha$. System \eqref{dtsyst} being controllable, the sufficient condition in this theorem is the application of an input that is PE of sufficiently large order as in~\autoref{defpe}. This excitation condition guarantees that the measured data is rich enough in information about the system dynamics and can be used, for example, for control design as described in the subsequent subsections. This fact has raised the question of whether persistence of excitation is a necessary condition for data-driven control. The answer to this question is in fact negative, and the study of sufficient and necessary conditions for data-driven system analysis and control has established the area of research of \textit{data informativity} \cite{Waardeetal2023}.

As described above, Willems' lemma was originally developed for discrete-time systems. In recent years, different approaches have been followed to obtain similar data-based representations for continuous-time systems. In \cite{LopezMuCDC2022}, an expression similar to \eqref{dtwillems} was used to express continuous-time input-output trajectories, but using time-varying Hankel matrices, and where the vector $\alpha$ satisfies a state-dependent differential equation. Since the use of state information is undesirable, in \cite{LopezMuRaLCSS2024} an alternative system representation was obtained by exploiting input-output trajectories and their derivatives. A different approach was followed in \cite{RapisardaCamWaa2023b}, where orthogonal bases are used to obtain approximate representations of the continuous-time trajectories. More recently, \cite{Schmitzetal2024} developed a system representation by computing Gramians of the measured data. Although we point out below some works that have used these results for data-based control of continuous-time systems, we focus most of the following exposition on the discrete-time case.

Before presenting data-driven control strategies, in the following subsection we briefly describe some of the extensions of the fundamental lemma that have been developed to allow its application to a wider class of dynamical systems or that relax the requirements on the data.

\subsection{Extensions to the Fundamental Lemma}

Willems' fundamental lemma requires the application of a persistently exciting input to a controllable system in order to obtain the data-based representation in \eqref{dtwillems}. However, in practice it may not always be possible to arbitrarily excite a system due to, for example, safety constraints. Moreover, this approach excludes considering uncontrollable systems that may be of interest in certain applications. In \cite{MarkovskyDor2023}, a \textit{generalized} persistence of excitation condition for the data was presented, consisting of the rank condition
\begin{equation}
	\text{rank} \left( \begin{bmatrix}
		H_L(u^d) \\ H_L(y^d)
	\end{bmatrix} \right) = n_\mathrm{u} L+n_\mathrm{x}
	\label{generalcond}
\end{equation}
with $L > l$, and where $l$ is the lag, or observability index, of \eqref{dtsyst}. It is shown in \cite{MarkovskyDor2023} that \eqref{dtwillems} is a data-based representation of system \eqref{dtsyst} if and only if \eqref{generalcond} holds.

Another important extension of~\autoref{thmdtwill} is relaxing the requirement of collecting a single, long enough PE trajectory from the system. In some applications, it may be inconvenient to collect long trajectories from \eqref{dtsyst}, for example if the open-loop system is unstable. Moreover, the user may face intermittent loss of data. In \cite{vanWaardePerCamTes2020}, the fundamental lemma is extended to the case where multiple trajectories are available. First, the concept of \textit{collective} persistence of excitation was presented by considering $q$ different input sequences, $u^{d,i} = \{ u_k^{d,i} \}_{k=0}^{N_i-1}$, $i=1,\ldots,q$, selected such that $\text{rank} \left( [H_L(u^{d,1}) \quad H_L(u^{d,2}) \; \cdots \; H_L(u^{d,q})] \right) = n_\mathrm{u} L$. To satisfy this rank condition, it is not necessary that any one of the input sequences is PE as in~\autoref{defpe}. Using the data collected from the system, the data-based representation now becomes
\begin{equation}
		\begin{bmatrix}
			H_L(u^{d,1}) & H_L(u^{d,2}) & \cdots & H_L(u^{d,q}) \\ H_L(y^{d,1}) & H_L(y^{d,2}) & \cdots & H_L(y^{d,q})
		\end{bmatrix} \alpha = \begin{bmatrix}
			\bar u \\ \bar y
		\end{bmatrix}.
\end{equation}

Besides the LTI systems described above, the fundamental lemma has also been extended to other classes of dynamical systems including linear time-varying systems \cite{NortmannMyl2021}, linear parameter-varying systems \cite{VerhoekTotHaeKoc2021}, descriptor systems \cite{SchmitzFaulWor2022}, and multiagent systems \cite{lopez2024databased}. Stochastic systems were considered in \cite{PanOuFau2023}, and representations from noisy data were described in \cite{YinIanSmi2021}. In \cite{AlsaltiMaLoMuTAC2024}, the case when only irregularly measured data is available is studied. Links between the fundamental lemma and kernel regression were investigated in \cite{MolodchykFaul2024}.

Data-based representations for classes of nonlinear systems have also been obtained \cite{RuedaEscobedoSch2020,Markovsky2023,BerberichAll2020,AlsaltiLoBeAlMuTAC2023,ShangCorZhe2024}. These results often exploit knowledge about the nonlinear terms in the system model, for example in the form of a dictionary of basis functions, and then they take advantage of the known results for linear systems. For example, Hankel matrices are used as in \eqref{dtwillems}, but with nonlinearities in their entries.

\subsection{Data-Based Control Design}
In the following subsections, we briefly describe some basic approaches for direct data-driven control in the existing literature. Many more works and extensions than those presented here can be found in the surveys \cite{MarkovskyDor2021,Waardeetal2023} as well as the book \cite{WaardeCamlibelTrentelman2025}.

Most of the control design methods in the literature consider state-feedback controllers. Hence, we begin by considering a linear system represented only by the equation \eqref{dtsysx}, such that the state $x$ can be measured. Later, we describe some results that do not require this assumption. Hence, assume throughout this subsection that system \eqref{dtsysx} is controllable. After collecting $N+1$ samples of input and state data, $\{ u_k^d \}_{k=0}^N$, $\{ x_k^d \}_{k=0}^N$, define the matrices $U$, $X$, and $X^+$ in \eqref{eq:data_u}, \eqref{eq:data_x}, and \eqref{eq:data_x+}, respectively. Notice that, by definition, the relation $X^+=AX+BU$ holds. These matrices were used in \cite{DePersisTes2020} to formulate a simple data-based method to design stabilizing controllers using the following linear matrix inequality (LMI)
\begin{equation}
	\begin{bmatrix}
		X \Theta & X^+ \Theta \\  (X^+ \Theta)^\top & X \Theta
	\end{bmatrix} \succ 0.
	\label{lmicontrol}
\end{equation}
The objective in \eqref{lmicontrol} is to find any $\Theta \in \mathbb{R}^{N \times n_\mathrm{x}}$ that satisfies the LMI. Using any such matrix, a state-feedback gain can be computed as
\begin{equation}
	K = U \Theta \left(X \Theta \right)^{-1}.
	\label{stabilizingk}
\end{equation}
It is shown in \cite{DePersisTes2020} that, if the data satisfy the rank condition
\begin{equation}
	\text{rank} \left( \begin{bmatrix} U \\ X \end{bmatrix} \right) = n_\mathrm{u}+n_\mathrm{x},
	\label{rankcontrol}
\end{equation}
then the control input $u=Kx$ with $K$ as in \eqref{stabilizingk}, and where $\Theta$ satisfies \eqref{lmicontrol}, stabilizes the system \eqref{dtsysx}. That is, the closed-loop matrix $A+BK$ is Schur stable, meaning that all its eigenvalues are inside the unit circle. Conversely, any stabilizing feedback gain $K$ for system \eqref{dtsysx} can be written as \eqref{stabilizingk}, where $\Theta$ satisfies \eqref{lmicontrol}. These facts can be seen by noticing that \eqref{lmicontrol} implies that $X \Theta$ is symmetric and positive definite, and hence we can define $P:= X \Theta$. Moreover, notice that $X^+ \Theta = (AX + BU) \Theta = (A+BK) P$, where we used $U \Theta = K X \Theta$ from \eqref{stabilizingk}. By the Schur complement, \eqref{lmicontrol} also implies $X \Theta - X^+ \Theta (X \Theta)^{-1} (X^+ \Theta)^\top  \succ 0$, which yields the Lyapunov condition $(A+BK) P (A+BK)^\top -P \prec 0$ that shows stability. Note that computing a stabilizing controller from \eqref{lmicontrol}-\eqref{stabilizingk} does not require any knowledge of the system model. The condition in \eqref{rankcontrol} can be guaranteed by applying a PE input of order $n_\mathrm{x}+1$ to the system, according to Willems' fundamental lemma \cite{willems_lemma,DePersisTes2020}. 

More sophisticated data-based control methods have been developed, considering requirements for robustness, optimality, and other performance characteristics. The case in which the measured data is corrupted by noise has been addressed with the development of robust controllers \cite{BisoffiPerTes2022,Waardeetal2022,BerberichScheAll2023}. Different approaches to solve the data-based optimal control problem have been investigated \cite{DePersisTes2020,DorflerTesDe2021,LopezAlsMue2023}. A model-reference controller was proposed in \cite{breschi2021direct} for the system to emulate the behavior of a desired model. The pole placement problem was considered in \cite{Bianchin2023,BisoffiDePerTes2023}. The data-based control of continuous-time systems has also been investigated \cite{RapisardaWaaCam2024,EisingCor2025,LopezMul2026,Possieri2026}. Of current interest is the extension of these methods to (classes of) nonlinear systems. In this regard, diverse results have already been obtained using different approaches \cite{DePersisTesi2023,HuPerTes2025,YuanCor2022,VerhoekKoeHaeTot2023,MartinSchAll2023,AlsaltiDepLopMul2025}.

We finally comment on control design results that use only input-output data, and no state information is required. Already in \cite{DePersisTes2020} it was noted that an alternative, nonminimal state for a linear system of the form \eqref{dtsyst} can be constructed using past measurements of input-output data as $\xi_k = [	u_{[k-n_\mathrm{x},k-1]}^\top \quad y_{[k-n_\mathrm{x},k-1]}^\top ]^\top \in \mathbb{R}^{(n_\mathrm{u}+n_\mathrm{y}) n_\mathrm{x}}$. Although $\xi_k$ is a valid state of the system, its use for data-based control is limited to systems for which the relation $n_\mathrm{y} l = n_\mathrm{x}$ holds \cite{BerberichScheAll2023,AlsaltiLoMuTAC2025}. However, for general linear systems it holds that $n_\mathrm{y} l \geq n_\mathrm{x}$ and, if the strict equality does not hold, the rank condition (similar to \eqref{rankcontrol}) that is used to obtain guarantees for the controller is not satisfied. In \cite{AlsaltiLoMuTAC2025}, this problem is solved by instead defining the state
\begin{equation}
	z_k = \begin{bmatrix}
		u_{[k-l,k-1]} \\ \Gamma y_{[k-l,k-1]}
	\end{bmatrix} \in \mathbb{R}^{n_\mathrm{u} l + n_\mathrm{x}},
\end{equation}
where the matrix $\Gamma$ can be computed from input-output data. It is shown that $z_k$ is also a valid state of system \eqref{dtsyst}, and it can be used for control design of general controllable linear systems. Indeed, any of the data-based strategies that have been developed for state-feedback control can be used to obtain a (dynamic) output-feedback controller of the form $u_k = K z_k$.  Other methods for data-based output-feedback control have also been proposed~\cite{WaardeEisetal2024,LiBisDepMon2026}. Besides control design, data-driven approaches for the dual problem of state estimation have been developed as well~\cite{TuranFer2022,ShiLiaJon2022,Liuetal2024,WolffLopMulTAC2024}.

\subsection{Data-Driven Predictive Control}
One particular direct data-driven control method that has received significant attention in the literature is data-driven predictive control (DDPC).
In DDPC, at each time $t$, given the past $T_{\mathrm{ini}}$ inputs $u_{[t-T_{\mathrm{ini}},t-1]}$ and outputs $y_{[t-T_{\mathrm{ini}},t-1]}$ with $T_{\mathrm{ini}} \geq l$, the following optimization problem is solved:
\begin{subequations} \label{DDPC_problem}
    \begin{align}
    \underset{\bar{u}(t),\bar{y}(t),\alpha(t)}{\text{minimize}}
    &\sum_{k=0}^{L-1}\ell(\bar{u}_k(t),\bar{y}_k(t)) \label{cost_DDPC} \\
    \text{s.t.} \quad &\begin{bmatrix}
        H_{L+T_{\mathrm{ini}}}(u^d) \\
        H_{L+T_{\mathrm{ini}}}(y^d)
    \end{bmatrix}\alpha(t) = \begin{bmatrix}
        \bar{u}(t) \\
        \bar{y}(t)
    \end{bmatrix} \label{prediction_model_DDPC} \\
    &\begin{bmatrix}
        \bar{u}_{[-T_{\mathrm{ini}},-1]}(t) \\
        \bar{y}_{[-T_{\mathrm{ini}},-1]}(t)
    \end{bmatrix} = \begin{bmatrix}
        u_{[t-T_{\mathrm{ini}},t-1]} \\
        y_{[t-T_{\mathrm{ini}},t-1]}
    \end{bmatrix}  \label{initial_condition_DDPC} \\
    &\bar{u}_k(t)\in\mathbb{U},~\bar{y}_k(t)\in\mathbb{Y},~0\leq k \leq L-1. \label{input_output_constraints_DDPC}
\end{align}
\end{subequations}
In problem~\eqref{DDPC_problem}, the decision variables $\bar{u}(t)$ and $\bar{y}(t)$ correspond to the predicted input and output sequences, predicted at time~$t$.
These are of length $L+T_{\mathrm{ini}}$, and the first $T_{\mathrm{ini}}$ elements of these sequences are fixed to the past $T_{\mathrm{ini}}$ inputs and outputs via the constraint~\eqref{initial_condition_DDPC}.
This ensures that the predicted future output sequence is correctly initialized according to the current state of the system. Constraint~\eqref{prediction_model_DDPC} replaces the system model that is typically used in classical, model-based MPC and uses the data-driven description~\eqref{dtwillems} from the fundamental lemma instead.
Moreover,~\eqref{input_output_constraints_DDPC} specifies the input and output constraints that the system should satisfy.
Finally, the stage cost~$\ell$ in~\eqref{cost_DDPC} is typically chosen as the classical quadratic cost function in the DDPC literature, i.e.,
\begin{align}
    \ell(u,y)=(u-u^s)^\top R(u-u^s)+(y-y^s)^\top Q(y-y^s) \nonumber
\end{align}
with $R>0$ and $Q\geq 0$ and $(u^s,y^s)$ being a desired input/output setpoint to be stabilized, but also other cost functions can be used depending on the desired control objective.
As usual in predictive control, problem~\eqref{DDPC_problem} is solved and then the first element of the optimal input sequence~$\bar{u}^*(t)$ is applied to the system, i.e., $u_t:=\bar{u}_0(t)$, before the procedure is repeated at the next time instant $t+1$.

DDPC based on problem~\eqref{DDPC_problem} has been introduced under the name \acitalic{DeePC} in~\cite{CoulsonLygDoe2019} and since then, many properties, extensions and related DDPC variants have been studied in the literature. 
In case of noisy measurements (both offline and online), one cannot exactly predict future trajectories via~\eqref{prediction_model_DDPC} and fixing the initial condition via~\eqref{initial_condition_DDPC} might be inconsistent.
To this end, one typically employs additional slack variables together with suitable regularization terms in~\eqref{DDPC_problem}. 
In particular, $\bar{y}(t)$ in~\eqref{prediction_model_DDPC} is replaced by $\bar{y}(t)+\sigma(t)$, where the slack variable $\sigma(t)$ ensures that the constraints~\eqref{prediction_model_DDPC}--\eqref{initial_condition_DDPC} are feasible despite noisy data.
Moreover, regularization terms $g_1(\sigma(t))$ and $g_2(\alpha(t))$ are added to the cost function~\eqref{cost_DDPC}.
Typical choices are $\ell_1$ and $\ell_2$ regularizations, as proposed, e.g., in~\cite{CoulsonLygDoe2019,BerberichKoeMueAll2021}, but also other choices have been used.
These regularization terms robustify the DDPC scheme and avoid overfitting to the noisy data.
A detailed study of different regularizations as well as further interpretations of their effects can be found in~\cite{coulson2019regularized,DorflerCoulMarTAC23,BreschiChiFor2023,KlaedtkeSchulzeDarupat25}.
Further stochastic DDPC schemes for settings including disturbances in the dynamics and noisy data have been proposed in, e.g., \cite{PanOuFaul2025,Kerzetal2023,Yin2024stochastic}.

Another aspect that has been studied in the DDPC literature is sample and computational efficiency. 
Namely, the standard formulation~\eqref{DDPC_problem} is such that the required minimum number of offline data points $u^d$ and $y^d$ is given by $N\ge(n_\mathrm{u}+1)(L+T_{\mathrm{ini}}+n_\mathrm{x})-1$ in order to satisfy the PE condition in~\autoref{thmdtwill}, scaling linearly with the prediction horizon $L$. Moreover, the dimension of $\alpha(t)$ scales linearly with the number $N$ of collected  data points. 
To address these issues, different DDPC variants have been proposed that modify the data-driven prediction model~\eqref{prediction_model_DDPC} to reduce the amount of required offline data and/or the number of optimization variables of problem~\eqref{DDPC_problem}, see, e.g., \cite{ODwyerTCST23,ZhangCSL23,Alsalti_eDDPC_26}.

In order to study closed-loop properties of DDPC such as stability, safety, and robustness, one can conceptually use the same tools as in classical model-based MPC, in particular adding suitable terminal constraints or using a long enough prediction horizon $L$. 
On the other hand, the details in deriving these closed-loop properties are quite different than in model-based MPC, in particular in case of noisy (offline and online) data where the prediction model~\eqref{prediction_model_DDPC} is not exact anymore.
The first results on closed-loop stability and robustness have appeared in~\cite{BerberichKoeMueAll2021}; a recent review on available methods and systems-theoretic guarantees in DDPC can be found in~\cite{BerberichAll2025}.

One major challenge and currently very active line of research is to develop DDPC approaches for nonlinear systems.
Existing works exploit extensions of Willems' lemma for specific system classes and/or knowledge of a set of suitable basis functions such as~\cite{Lian_Jones_21,Alsalti_IFAC23,Lazar_ECC24,Verhoek_et_al_TAC26}, or use online data in the Hankel matrices such that the prediction model~\eqref{prediction_model_DDPC} serves as a local affine approximation of the nonlinear system~\cite{Berberich_nonlinearDDPC_TAC22}.\\

\begin{tutorialexample*}[label={ex:deepc}]
    {Controlling without a model}
    {http://tiny.cc/cdc-tut-ex4}
  Willems' fundamental lemma allows us to work with collected data instead of a parametric model. In this didactic example, this is illustrated by using the standard DDPC algorithm based on~\eqref{DDPC_problem} and examining how data length and regularization affect closed-loop performance in non-ideal settings. However, note that as explained above, other DDPC variants have been developed to specifically address nonlinear systems. The implemented framework can be found in 
  \cite{cfr-notebook} (\href{http://tiny.cc/cdc-tut-ex4}{DeePC}).

 We consider the terrain setup as in~\autoref{ex:model-learning} and~\autoref{ex:gp-mpc} and assume that we have access to position measurements only (i.e., $y=p$). We solve the goal-reaching problem with $\ell(u,y)=(y-p^\mathrm{s})^2+0.02\,u^2$, $L=20$ and $T_\mathrm{ini}=4$. To collect the offline data, a uniformly random input of length~$N$ is applied. 

In the ideal flat-terrain case, the system is \ac{LTI}, with the data-length bound $N\ge 51$ required to satisfy the PE condition from~\autoref{thmdtwill}. \autoref{fig:tutorial-deepc-flat} shows that \textit{(i)} with a much smaller dataset, problem~\eqref{DDPC_problem} is infeasible, and \textit{(ii)} once the bound is reached, the closed-loop behaviour coincides with that of \ac{MPC} using the exact model. Further increasing the data length beyond the theoretic bound does not result in better performance.

  \begin{tutorialfigure}
      \includegraphics[width=\linewidth]
          {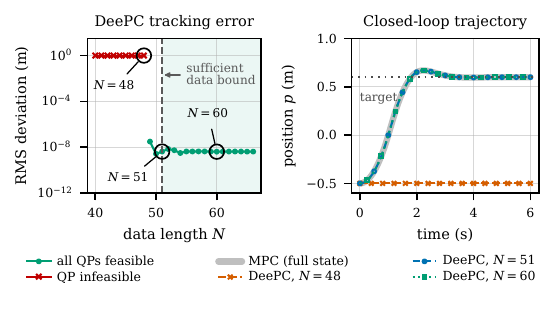}
      \captionof{figure}{
          DeePC on flat terrain, where \ac{LTI} assumption holds. Left: Deviation of the DeePC controller from the full-state \ac{MPC} versus data length~$N$, where the shaded region corresponds to  guaranteed \ac{PE}. 
          Right: trajectories for the circled~$N$ cases.
      }
      \label{fig:tutorial-deepc-flat}
  \end{tutorialfigure}

We then consider bumpy terrain with measurement noise, where the \ac{LTI} assumption no longer holds. The top panels of~\autoref{fig:tutorial-deepc-bumpy} show that more data are needed to stabilize the system, while a nonzero error remains due to the non-ideal setting. The bottom panels of~\autoref{fig:tutorial-deepc-bumpy} show how regularization balances fitting the measured history against biasing the predictor; more information about the effects and different types of regularization can be found in \cite{coulson2019regularized,DorflerCoulMarTAC23,BreschiChiFor2023,KlaedtkeSchulzeDarupat25}. With suitable tuning, \ac{DeePC} approaches the performance of \ac{MPC} based on the imperfect flat-terrain model. 

  \begin{tutorialfigure}
    \includegraphics[width=\linewidth]
          {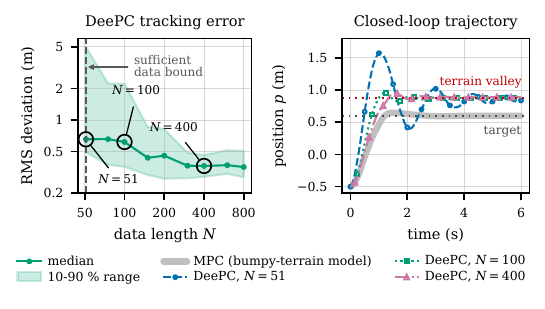}
    \includegraphics[width=\linewidth]
          {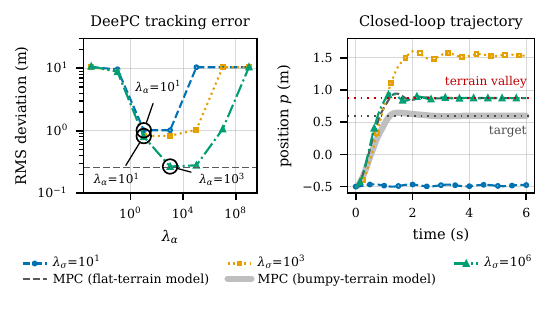}
      \captionof{figure}{
          \ac{DeePC} with bumpy terrain and measurement noise. Top-Left: Deviation from the full-state \ac{MPC} (median and 10-90\,\% range computed over ten trials). Top-Right: trajectories for the circled~$N$ cases. Bottom-Left: Deviation from the full-state \ac{MPC} with the bumpy-terrain model for different regularization parameters $\lambda_\alpha$ and $\lambda_\sigma$, against the \ac{MPC} with the flat-terrain model. Bottom-Right: trajectories for the circled settings.
      }
      \label{fig:tutorial-deepc-bumpy}
  \end{tutorialfigure}
  \end{tutorialexample*}
\section{Reinforcement Learning, Imitation Learning, and Bayesian Optimization}
\label{sec:rl}
This section focuses on learning-based control methods that directly learn a policy~\eqref{eq:policy} from data, often through iterative interaction with the environment, without first identifying a model of the system dynamics (\autoref{sec:dynamics}) or leveraging data-driven control techniques building on Willems' fundamental lemma (\autoref{sec:ddc}). Typically, these methods employ a parametrized policy $\pi_\theta(\cdot)$, where the parameters $\theta$ may belong to a function approximator (often a neural network) or to a classical controller structure such as PID, LQR, or MPC. At a high level, learning a policy~\eqref{eq:policy} to optimize performance~\eqref{eq:objective} can then be formulated as a parametric optimization problem
\begin{equation}
    \min_\theta J(\theta).
    \label{eq:general_optimization_directlearning}
\end{equation}
Depending on the method and the available prior knowledge, $\theta$ 
can differ substantially in both dimension and interpretation. When 
$\theta$ are the parameters of a high-capacity function approximator 
such as a neural network, \eqref{eq:general_optimization_directlearning} 
represents a general, expressive problem formulation, capable of 
representing a wide class of policies without imposing strong structural 
assumptions on $\pi$. For this class of problems, we focus on 
state-of-the-art \ac{RL} and imitation learning 
approaches. In particular, we discuss how to address key challenges 
for real-world deployment, including safety, robustness, stability, 
and sample efficiency.

When instead specifying a fixed controller structure as named above, 
prior knowledge about the system or desired closed-loop behaviour is 
typically built directly into~$\pi$, and $\theta$ is correspondingly 
lower dimensional and often (more) interpretable. In this case, 
problem~\eqref{eq:general_optimization_directlearning} is also 
referred to as \emph{controller tuning}. Here, we focus on Bayesian 
optimization methods for safely optimizing controller parameters 
directly from input-output data collected from the system.

\autoref{tab:method-comparison} (bottom) provides an overview of the methods discussed in this section and their selected main characteristics.

\subsection{Model-Free Reinforcement Learning}
\ac{RL} studies how an agent can learn to make goal-directed sequential decisions through interaction with an environment.
In the context of the learning-based control problem from~\autoref{sec:problem_formulation}, the agent can be viewed as an adaptive controller that learns a policy \eqref{eq:policy} from the interaction data, which then determines the control action. The environment encompasses the controlled system together with all relevant external factors.
Formally, the environment is often modeled as a Markov decision process (MDP) $(\mathcal{S}, \mathcal{A}, p, r, \gamma)$, where $\mathcal{S}$ is the state space, $\mathcal{A}$ the action space, $p(x_{k+1}\:|\:x_k,u_k)$ the transition dynamics, $r(x_k,u_k)$ the reward function, and $\gamma \in [0,1)$ a discount factor. 
Even though the notation differs slightly and we maximize rewards instead of minimizing costs, the MDP formulation and the state-space formulation \eqref{eq:system} are equivalent. 
The MDP transition kernel $p(x_{k+1}\:|\:x_k,u_k)$ is the conditional distribution over successor states induced by the dynamics $f$ and disturbances $w_k$. 
The goal is to learn a policy $\pi(u_k|x_k)$ that maximizes the expected cumulative reward
\begin{equation}
    \max_\pi \mathbb{E}_\pi\left[\sum_{k=0}^\infty \gamma^k r(x_k,u_k)\right],
    \label{eq:RLobjective}
\end{equation}
\edits{where $\pi$ is a conditional distribution and $\mathbb{E}_\pi$ denotes the expectation under the policy $\pi$ for generating actions.}

Classical RL approaches are mostly focused on tabular settings, i.e., with finite and discrete state and action spaces. For control-oriented tasks, we require continuous state-action spaces and thus, focus on modern deep RL, where the policy $\pi$ is usually represented by a deep neural network, which leads to the notation $\pi_\theta$, where $\theta$ are the weights of the neural network. 
We can now identify our RL objective \eqref{eq:RLobjective} with the general formulation \eqref{eq:general_optimization_directlearning}:
\begin{equation}
    \min_\theta J(\theta) = -\max_\theta \mathbb{E}_{\pi_\theta}\left[\sum_{k=0}^\infty \gamma^k r(x_k,u_k)\right].
\end{equation}

The basic idea of RL is to learn a competent control policy by 
interacting with the environment: repeatedly trying out control 
inputs (actions) and observing the resulting state transitions and 
incurred costs (rewards). The data is thus typically composed of 
state~\eqref{eq:data_x}, input~\eqref{eq:data_u}, and reward data. 
Different categories of RL emerge depending on what objects are 
learned and how the search for an optimal policy is structured.
We first consider model-free RL, i.e., methods that do not construct 
a model of the system dynamics as part of solving the RL problem. 
The main categories within model-free RL are \textit{(i)}~policy-based, 
\textit{(ii)}~value-function-based, and \textit{(iii)}~actor-critic-based approaches, which we discuss next. 
Afterwards, we discuss why learning a transition model 
alongside the policy can nonetheless be beneficial for~RL.

Policy-based approaches are the closest to our optimization objective \eqref{eq:RLobjective} and yield iterative updates of the type
\begin{equation}
    \theta_{k+1}= \theta_k - \alpha \widehat{\nabla J(\theta_k)},
\end{equation}
where $\alpha$ is the learning rate or step size and $\widehat{\nabla J(\theta_k)}$ is an estimate of the gradient of $J$ with respect to $\theta$. Clearly, the challenge lies in estimating $\widehat{\nabla J(\theta_k)}$ from data. 
The policy gradient theorem provides the theoretical basis for this 
estimation: it expresses the gradient of the expected return as an 
expectation over trajectories sampled from the current policy, 
involving only the gradients of the log-policy (i.e., the log 
probabilities of the chosen actions) rather than the unknown system 
dynamics, thus allowing the gradient to be estimated purely from 
sampled data.
A didactic overview including the canonical policy gradient algorithm REINFORCE is provided in \cite[Sec.13]{sutton2018reinforcement}.

Value-based methods center around the action value function
\begin{equation}
    Q_\pi(x_k,u_k) = \mathbb{E}_\pi\left[ \sum_{l=0}^\infty \gamma^l \edits{r(x_{k+l}, u_{k+l})}\:\:\middle|\:\: x_k, u_k \right].
\end{equation}
Intuitively, the $Q$-function quantifies the value of state-action pairs. 
The size of the action space is usually very limited. The famous \ac{DQN} paper~\cite{mnih2015human} varies the valid actions between 4 and 18 depending on the specific Atari game that is considered. \edits{The optimal policy can be obtained via $\pi^*(x_k) = \argmax_u Q^*(x_k,u)$.}
Modern methods learn the $Q$-function via deep learning approaches to scale to large state-spaces. 
Accordingly, in pure value-based methods, we consider the parameters $\theta_Q$ that are used to learn the $Q$-function and not a policy.

Actor-critic methods combine policy gradients with value-based methods, leveraging strengths of both worlds. They typically maintain an estimate of the policy $\pi$ (the \emph{actor}) with parameters $\theta$ and a value function (the \emph{critic}) with other parameters $\theta_Q$. 
Intuitively, the value estimate leads to normalized learning signals that usually result in more stable and efficient algorithms than pure policy-gradient methods.
Notable examples include \ac{A3C} \cite{mnih2016asynchronous}, \ac{TRPO} \cite{schulman2015trust}, \ac{PPO} \cite{schulman2017proximal}, and \ac{TD3} \cite{fujimoto2018addressing}. 

Actor-critic methods are among the model-free RL approaches that are most suitable for control problems and require very little prior knowledge. At the same time, they are typically extremely data hungry and come without any safety guarantees. \edits{Data efficiency is a main motivation for \emph{model-based \ac{RL}}, which is discussed next.}

\edits{The above mentioned \ac{RL} approaches generally assume that the learning agent can actively sample the environment, i.e., query it by applying actions and observing the resulting next state and reward. In this sense, they are sometimes referred to as \emph{online} RL. The environment may be either a simulator or a physical system, such as a robot. In robotics, a common approach is to learn a policy in simulation during the design phase and subsequently deploy it on physical hardware, which often requires techniques such as domain randomization and sim-to-real adaptation to bridge the gap between simulated and real environments~\cite{muratore2022robot}. Some approaches instead learn directly from hardware interaction~\cite{bloesch2022towards} or even adapt the policy during deployment~\cite{hansen2021deployment}. In contrast, \emph{offline \ac{RL}} considers the setting in which the agent learns solely from a fixed dataset of previously collected interactions, without actively querying the environment. This is particularly relevant when interaction with the environment is undesirable or impractical, as it enables learning from historical data alone. A key challenge in offline RL is distribution shift: the learned policy may encounter state-action pairs that are poorly represented in the fixed dataset~\cite{fujimoto2019offpolicy}.} \\

\begin{tutorialexample*}[label={ex:rl},
    title={\tutorialtitlelabel~\thetcbcounter
           \titlecodelink[Code 1]{https://github.com/Data-Science-in-Mechanical-Engineering/RLLBC}%
           \titlecodelink*[Code 2]{http://tiny.cc/cdc-tut-ex5-saferl}}]
  {Reinforcement learning}{}
  We consider the standard cart-pole system, where the objective is to swing the pole into the upright position and balance it there. We use a DQN agent~\cite{mnih2015human}, adapted from~\cite{rllbc_paper} (\href{https://github.com/Data-Science-in-Mechanical-Engineering/RLLBC/blob/main/deep_examples/reinforce.ipynb}{REINFORCE on CartPole}). 
  In RL, the control objective is formalized as the reward. In this case, the reward function is defined as $r(x) = \frac{1 + \cos\theta}{2}\cdot (1 - (\frac{p}{p_{\text{bound}}})^{2}) \cdot \frac{1 + 10^{-(\dot\theta/5)^{2}}}{2},$ where $p$ denotes the cart position, $\theta$ the pole angle, and $\dot{\theta}$ the angular velocity. That is, the reward expresses the control goal of swinging the pole upright, keeping the cart within its position bounds, and minimizing angular velocity to stabilize the pole. An episode terminates when the cart position leaves the interval $[-p_{\text{bound}},p_{\text{bound}}]$. The control input is discrete and consists of applying a fixed force to the left, applying a fixed force to the right, or applying no force.

\autoref{fig:tutorial-RL} illustrates several characteristic properties of \ac{RL} applied to a control task. After training, the learned policy achieves a fast swing-up and successfully balances the pole near the upright position. In contrast to supervised learning, the performance during training is highly stochastic and can temporarily deteriorate. This is partly a consequence of the $\epsilon$-greedy exploration strategy used by the DQN agent: with probability $\epsilon$, the agent selects a random action rather than the action predicted by the current policy. Although such exploratory actions can intermediately lead to poor trajectories and low rewards, exploration is essential for discovering potentially better control actions.
A further characteristic of \ac{RL} is its reliance on repeated interactions with the environment. The agent must collect a large number of transitions through trial and error before it can learn a good approximation of the $Q$-function. The neural network parameters are subsequently optimized based on these interactions.

DQN~\cite{mnih2015human} was one of the first highly successful deep RL algorithms, though limited to discrete actions, and it triggered the development of many follow-up methods, including approaches for continuous action spaces.  
Further implementations of tabular and deep RL algorithms can be found in~\cite{rllbc_paper}.
Additional material and a comparison of \ac{PPO} and \ac{MPC} on a mountain car example can be found in~\cite{cfr-notebook} (\href{http://tiny.cc/cdc-tut-ex5-ppo}{PPO}).

\edits{
Although specific reward design can encourage safe behaviour, standard RL does not inherently guarantee satisfaction of constraints.
Safety filters can address this limitation by minimally modifying the agent's potentially unsafe actions to enable safe exploration during training~\cite{wabersich2021predictive,bejarano2024safety,yang2026cbfrlsafetyfilteringreinforcement,eisele2026dyna,frauenknecht2026uncertainty}.
To illustrate this approach, we impose the cart-position constraint $p \in [-1,\,2.4]$ and incorporate a CBF-based safety filter~\cite{yang2026cbfrlsafetyfilteringreinforcement} into the DQN agent's interactions with the environment throughout training.
An example implementation of the safety filter is provided in~\cite{cfr-notebook}~(\href{http://tiny.cc/cdc-tut-ex5-saferl}{Safe RL}).
As shown in~\autoref{fig:tutorial-SafeRL}, the filtered agent satisfies the constraint and achieves high episode returns with fewer environment interactions, illustrating how incorporating safety into exploration can also improve learning efficiency.
}

  \begin{tutorialfigure}
      \includegraphics[width=\linewidth]
          {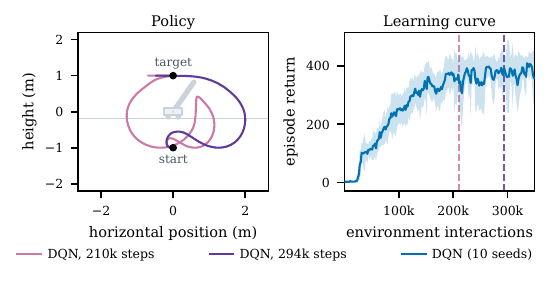}
      \captionof{figure}{
          Example behaviour of a DQN agent learning to swing up and balance a pole on a cart-pole system. The left plot shows two trajectories at different stages of the learning process, while the right plot shows the corresponding learning curve. The solid line represents the mean performance across runs, while the shaded area indicates the corresponding standard deviation. 
      }
      \label{fig:tutorial-RL}
  \end{tutorialfigure}

  \begin{tutorialfigure}
      \includegraphics[width=\linewidth]
          {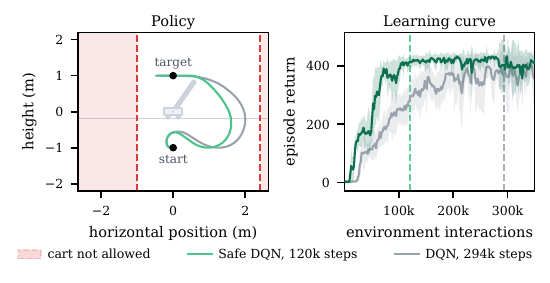}
      \captionof{figure}{
          \edits{Comparison of DQN with safety filter (green) and vanilla DQN (grey). The left plot shows representative pole-tip trajectories from the two agents, with the red shaded regions indicating forbidden cart positions. The right plot shows the corresponding learning curves. In this example, the safety filter enforces the safety constraints while improving learning efficiency, enabling the agent to achieve high episode returns with fewer environment interactions.}
      }
      \label{fig:tutorial-SafeRL}
  \end{tutorialfigure}
  
\end{tutorialexample*}

\subsection{Model-Based Reinforcement Learning}
At the core of model-based RL is a dynamics model that predicts future state transitions and enables planning and policy improvement.
This model $\hat{p}_{\theta_M}(x_{k+1}|x_k,u_k)$ is typically learned from data and does not need to be known a priori. 
Clearly, this adds the parameters $\theta_M$ to the overall optimization problem.

Next, we distinguish between two major lines of research in model-based RL: \textit{(i)} Dyna-style or background planning and \textit{(ii)} decision-time planning. 
Dyna-style methods primarily use the model to improve learning efficiency by generating additional training data. Modern methods utilize ensembles of probabilistic networks to learn the transition probabilities \cite{janner2019trust} and consider model uncertainty for improved rollouts \cite{frauenknecht2024trust,frauenknechtrollouts}.
Decision-time planning methods leverage the model to improve action selection through online look-ahead search. Conceptually, decision-time planning can be viewed as an instance of \ac{MPC} \cite{chua2018deep}.

While both approaches can yield gains in sample efficiency, the model serves fundamentally different roles: in the former, it supports policy learning, while in the latter, it directly guides decision making on the next action $u_k$.
In traditional control, dynamics models are typically grounded in physical principles and used to establish theoretical guarantees, such as stability and safety. In model-based RL, by contrast, models are primarily learned or employed to improve sample efficiency. Despite these distinct motivations, the two paradigms exhibit strong conceptual and methodological overlap, enabling fruitful cross-pollination and the transfer of advances between the fields \cite{xue2025full,reiter2026synthesis}.

In particular, access to a model and the integration of control-theoretic techniques (see~\autoref{sec:safety}) provide a means to incorporate different types of safety guarantees in RL. Depending on the specific notion of safety considered, representative approaches include the use of soft constraints \cite{achiam2017constrained,as2025actsafe}, Lyapunov functions~\cite{berkenkamp2017safe}, barrier-function-based methods \cite{wang2023enforcing}, and safety filters that modify or restrict policy actions to ensure constraint satisfaction \cite{lavanakul2024safety,eisele2026dyna, frauenknecht2026uncertainty}.

For the above methods, we consider full state observations, i.e., the true state $x_k$ is assumed to be measured and provided to the policy.
In practice, however, one often feeds all available 
measurements into the policy without explicit knowledge of whether 
they constitute the full state, and may additionally include 
historical observations to capture temporal dependencies.
Beyond this, there is work that looks at latent state space models (sometimes also called \emph{world models}) considering output measurements \eqref{eq:measurement}. This type of approach is particularly relevant when acting on image data and gave rise to the Dreamer architecture \cite{hafnerdream} and other influential methods such as \ac{TDMPC}\cite{hansen2022temporal}. 
Safety remains an important consideration in this setting as well~\cite{huang2024safedreamer}.

\subsection{Imitation Learning}
The previous approaches focus on discovering optimal behaviour through trial-and-error via the reward signal. Instead, we can learn a policy by observing and mimicking an expert's behaviour, which results in a supervised learning problem. 

This expert could, for example, be a human. 
Behaviour cloning and inverse RL 
 form two major classes of imitation learning methods \cite{osa2018algorithmic}.
In behaviour cloning, we leverage a dataset of state-action pairs $\mathcal{D}=\{ (x_k,u_k) \}$, sometimes also with an additional context variable. We cast the problem as a supervised learning problem and approximate the policy $\pi$ that has created the control inputs via a neural network $\pi_\theta$. Inverse RL, on the other hand, learns the underlying reward function, and we obtain $R_{\theta_R}$, which can afterwards be used to find an optimal policy.  
Current active topics, especially in robotics, are large behaviour models \cite{barreiros2026careful}, vision-language-action models such as $\pi_0$ \cite{black2024pi_0}, and foundation models in general. A core component of training such models is imitation learning, often in combination with RL. The number of parameters of the underlying transformer models ranges easily in the magnitude of billions.

As an alternative to learning from humans, the expert could also be a machine or an expensive-to-evaluate policy, such as a sophisticated modern MPC.
Essentially, the MPC is used to create state-input data offline, which can be used as training data for a neural network to approximate the MPC via supervised learning.
\edits{The imitation learning policy is often much faster to evaluate during deployment than the original MPC and with constant runtime, which are core motivations for this approach in practice.}
This approach of approximate MPC (see \cite{gonzalez2023neural} for a recent survey) gives rise to research questions on constraint satisfaction despite approximation errors \cite{hertneck2018learning,hose2025approximate,zhang2020near,bonzanini2020toward, karg2020efficient}, jointly optimizing the MPC and neural network policy \cite{drgovna2022differentiable, cao2020deep}, how to approximate set-valued solution maps with generative models \cite{romer2025diffusion,julbe2025diffusion, okamoto2024deep, reske2021imitation}, adapting policies to parameter changes \cite{hose2024parameter, iqbal2025zero}, or even learning iterative solvers \cite{luken2024learning}, to name just a few.
Despite great theoretical efforts, there are only a few applications in hardware, for example, in power electronics \cite{lucia2020deep}, robotics \cite{nubert2020safe, reske2021imitation, julbe2025diffusion, hose2025mini}, or automotive production-grade electronic control units \cite{zhang2020near}.

\subsection{Controller Tuning with Bayesian Optimization}
In many practical applications, established controller structures such as PID controllers, LQR, or MPC with predefined cost formulations are often preferred due to their interpretability, robustness, or ease of implementation. However, even when the controller structure is fixed, its parameters must still be selected to achieve good performance on the actual system. This process is commonly referred to as controller tuning. At a high level, controller tuning coincides with optimization problem \eqref{eq:general_optimization_directlearning}, where the goal is to identify parameter values that optimize a given performance objective. A key distinction, however, is that the parameter space in controller tuning is typically much lower-dimensional than the policy parameter spaces considered in deep RL. In many cases, tuning involves only a few to several tens of parameters. This lower dimensionality enables the use of specialized optimization methods that often require substantially less data and may provide stronger theoretical or practical guarantees compared with general-purpose deep RL approaches.

Controller tuning is a long-standing problem and many methods exist, including PID 
auto-tuners~\cite{astrom2006advanced}, extremum seeking 
control~\cite{ariyur2003real}, virtual reference feedback 
control~\cite{campi2002virtual}, iterative feedback 
tuning~\cite{hjalmarsson2002iterative}, and evolutionary 
search~\cite{fleming2002evolutionary}, among many others. 
Here, we focus on \ac{BO}~\cite{garnett_bayesoptbook_2023}, 
which has gathered substantial momentum over the last decade and 
matured into a powerful framework for controller 
tuning~\cite{stenger2026decade, paulson2023tutorial}. 
\edits{Its sample efficiency and black-box formulation make it particularly attractive for tuning directly on physical systems, where evaluations can be costly and accurate simulation models may be unavailable.}

At the heart of BO for controller tuning lies the optimization problem \eqref{eq:general_optimization_directlearning}. We assume that the function $J$ is unknown, can be evaluated \edits{through closed-loop experiments}, and represents a control performance measure that depends on controller parameters $\theta$, which are typically low-dimensional. To guide the search for optimal parameters, BO constructs a probabilistic surrogate model of $J$, usually in the form of a Gaussian process,
$J \sim \mathcal{GP}(m(\theta), k(\theta, \theta'))$, where $m(\cdot)$ denotes the mean and $k(\cdot,\cdot)$ the covariance function (cf.~\autoref{subsec:probabilistic}). Alternatively, other probabilistic machine learning models can be used as surrogates, such as neural networks \cite{brunzema2025bayesian}. Based on the posterior predictions of the surrogate model, an acquisition function $\alpha(\theta, \mathcal{D})$, where $\mathcal{D}$ contains the past evaluations, determines which controller parameters are evaluated next. It quantifies the utility of candidate points by balancing exploration (e.g., testing regions with high uncertainty) and exploitation (i.e., refining solutions that are predicted to perform well).

Depending on the specific problem properties, many different approaches have been developed. Important research directions include controllers with higher-dimensional controller parameters~\cite{eriksson2019scalable, muller2021local}, safety during controller tuning~\cite{berkenkamp2016safeopt,berkenkamp2023bayesian,fiedlersafety, frohlich2022contextual, puigjaner2025performance}, and settings where changes in the system must be 
detected and the controller retuned accordingly~\cite{brunzemaevent,bogunovic2016time}. A broader overview can be found in \cite{stenger2026decade}.

\begin{tutorialexample*}[label={ex:bo}]
  {Bayesian optimization}{https://github.com/Data-Science-in-Mechanical-Engineering/tunecontrol}
  We consider a pole-balancing task adapted from~\cite{stenger2026decade} (\href{https://github.com/Data-Science-in-Mechanical-Engineering/tunecontrol/blob/main/examples/standard_bo_for_controller_tuning.ipynb}{CartPole}).
  For illustrative purposes, we tune only one parameter of a static feedback controller while keeping the remaining controller structure fixed. In \ac{BO}, we choose $J$ to capture the control objective, in this example, through a trajectory-averaged penalty on absolute state-tracking errors and control effort:
$J(\theta)=\frac{1}{N}\sum_{k=0}^{N-1}(10|e_k(\theta)|_1+|u_k(\theta)|)$, where $e_k=[x_k-r_k,\dot{x}_k,\phi_k,\dot{\phi}_k]^\top$ denotes the state-tracking error, $r_k$ the cart-position reference, and $u_k$ the control input.
That is, $J$ is evaluated from state and input trajectory data collected during a closed-loop experiment with a chosen gain $\theta$, using 30 seconds of data. BO seeks the gain that minimizes this cost.

In~\autoref{fig:tutorial-BO}, we illustrate the behaviour of a \ac{BO} algorithm using expected improvement as the acquisition function. Rather than evaluating the controller performance exhaustively, BO uses the available observations to construct a \ac{GP} surrogate model of the cost function (top plot) and selects the next controller parameter based on the acquisition function (bottom plot). 
The acquisition function therefore does more than simply search for the parameter with the highest uncertainty. It evaluates how promising each candidate is by considering both the predicted cost and the uncertainty of the prediction. This allows BO to focus evaluations on parameters that are likely to improve the current solution, while still exploring regions where the surrogate model is uncertain.
Each observation requires applying the corresponding controller parameter to the physical system and measuring the resulting cost~$J$.

By exploiting the prior control structure and tuning only a single parameter, the BO algorithm identifies a near-optimal parameter after only five iterations. This example highlights a key advantage of BO for controller tuning:  it can efficiently optimize expensive-to-evaluate objectives while  requiring only a small number of \tutorialcolumnbreak   system experiments.   While this example tunes a  single parameter, standard BO is typically well-suited to tuning a modest number of parameters. Advanced BO variants have been developed targeting specific desiderata of the tuning problem, such as safety considerations, high dimensionality, and time-varying objectives~\cite{stenger2026decade}.

  \begin{tutorialfigure}
      \includegraphics[width=\linewidth]
          {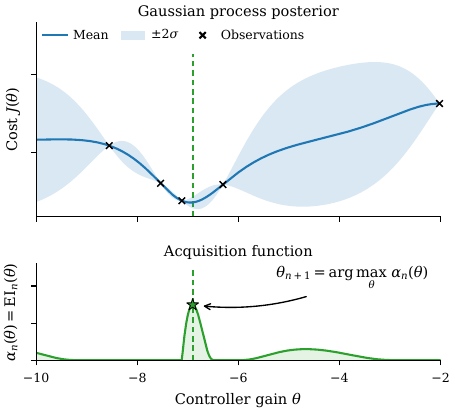}
      \captionof{figure}{
         Controller tuning with \ac{BO}. The top plot shows the \ac{GP} surrogate model of the cost function $J$, evaluated at different controller parameters $\theta$. The bottom plot shows the corresponding acquisition function (expected improvement), which quantifies the utility of evaluating each candidate controller parameter next.
      }
      \label{fig:tutorial-BO}
  \end{tutorialfigure}
\end{tutorialexample*}
\section{Applications}\label{sec:applicationsandbenchmark}
The developments reviewed in this tutorial have enabled learning-based control methods to be applied across a wide range of practical systems. Examples from the CDC Learning-Based Control Invited Sessions span healthcare, robotics and autonomous systems, energy and power systems, process and industrial control, and networked systems. Representative applications include adaptive medication tapering~\cite{gradu2023online}, autonomous driving and racing~\cite{bajcsy2019efficient, jain2020computing, menn2024lipschitz, 9484747}, quadrotor control~\cite{bansal2016learning,zhou2017design, zhou2019active,greeff2021exploiting}, manipulation~\cite{9683483, beckers2017stable,dohmann2021distributed},  multiagent coordination ~\cite{zhu2020trajectory,hock2016distributed,suriyarachchi2022multiagent,sorocky_share_2021,10433772},  power electronics~\cite{xu2024constraintsinformed} and energy management~\cite{degner2024adaptive}, process control~\cite{armenio2019model}, and wireless sensor networks~\cite{beuchert2020hierarchical}.
Across these domains, learning-based methods have been used to improve prediction under uncertainty, adapt controllers from data, reduce conservatism, and enhance performance while preserving desired closed-loop properties.

Robotics has been a particularly prominent application area. Robotic systems expose many of the central challenges in learning-based control, including nonlinear and uncertain dynamics, complex task specifications, limited data availability, real-time computation, perception and actuation uncertainty, and strict safety requirements. They therefore provide useful testbeds for  understanding the strengths and limitations of different methodological paradigms, as illustrated, for example, by existing benchmarks~\cite{yuan2022safecontrolgym,safelearn}.

Robotic applications also illustrate how learning-oriented and control-oriented approaches can complement one another in practice. \edits{Learning-oriented methods are flexible in two respects. They can optimize policies for complex sequential decision-making tasks, as in \ac{RL} for autonomous drone racing~\cite{kaufmann2023champion} and quadruped locomotion~\cite{rudin2022learning}, and they can model poorly known dynamics from data, as in learning terrain-dependent disturbance models for off-road driving~\cite{mckinnon2018experience}. Their performance on real systems, however, depends strongly on the coverage and quality of the training data. Robustification strategies such as domain randomization can compensate for part of the mismatch between training and deployment conditions, but the task performance sacrificed for this robustness is difficult to characterize in advance.} Control-oriented methods, in contrast, exploit prior knowledge of system dynamics and structure, which can improve data efficiency and support the analysis of closed-loop properties. For instance, differential flatness can substantially reduce the learning burden for quadrotor control~\cite{greeff2021exploiting} and imposing Lipschitz regularities informed by control theory can support formal closed-loop stability analysis~\cite{shi2019neural,zhou_bridging_2022}.

\edits{Recent work extends this complementarity along three axes: higher-dimensional systems, more expressive policy classes, and richer safety specifications. For instance, control-guided deep \ac{RL} has scaled from quadrotors, where safety filtering is incorporated into training~\cite{bejarano2024safety}, to humanoids, where \acp{CLF} shape rewards for dynamic locomotion~\cite{11346966}. Learned policies have likewise progressed from feedforward networks that approximate a single-valued \ac{MPC} law~\cite{hertneck2018learning,nubert2020safe} to diffusion models that represent multimodal distributions over \ac{MPC} solutions~\cite{julbe2025diffusion,romer2025diffusion}. Beyond more complex systems and more expressive policies, safety certifications are also expanding beyond geometric constraints such as collision avoidance to semantically-driven commonsense specifications required for unstructured, everyday scenarios~\cite{bahati2025risk,feng2025words,10933541}. Together, these developments illustrate a common direction: theory lends structure to learning as systems and policies grow more complex, while learning broadens the classes of systems, the range of tasks, and the types of safety specifications that  control can accommodate.}

\section{Conclusions and Outlook}
\label{sec:conclusionsandoutlook}
This tutorial paper reviewed key methodological directions in learning-based control that have emerged over the past decade, motivated by the CDC Learning-Based Control Invited Sessions. Across the surveyed approaches, a common theme is that machine
learning expands what can be modeled, optimized, and adapted,
while control theory provides structure for ensuring stability,
safety, robustness, and reliable closed-loop behaviour. As autonomous systems
move toward increasingly complex, uncertain and safety-critical real-world settings,  incorporating data-based methods directly into the closed loop will be critical for achieving reliable high-performance behaviour. 

We highlight several directions that we believe have the potential to drive further practical impact.
\begin{itemize}
\item \edits{\textit{Reliably scaling up learning-based control:}
In this tutorial, we touched upon families of methods that address how uncertainty can be accounted for in learned models and how policies can be optimized without explicit system identification, including through direct data-driven control,  \ac{RL}, \ac{IL}, and \ac{BO} (\autoref{tab:method-comparison}). 
There are several important directions we did not dive into details but are equally important for reliably scaling these methods to complex real-world problems. For instance, systems must be able to \textit{learn online and continually} as their dynamics, environments, and objectives change, while avoiding catastrophic forgetting, satisfying real-time computational and memory constraints, and preserving closed-loop guarantees. It is often also desirable to \textit{actively identify and explore poorly understood components}, while acquiring reliable   uncertainty estimates under continual updates. Finally, learning-based control can \textit{extend beyond the time-invariant, discrete-time systems} considered here to hybrid, time-varying, and distributed dynamics and possibly \textit{more tightly couple perception, state estimation, learning, and control} within a unified optimal decision-making framework. Addressing these challenges jointly will be essential for developing learning-based control systems that are capable and scalable without sacrificing reliability required for real-world deployment.}
\item \edits{\textit{Leveraging control for learning:} As discussed throughout this tutorial, learning-based control has expanded from structured approaches that integrate learned linear or nonlinear models into established control frameworks to direct data-driven methods for synthesizing policies, as well as flexible optimization regimes that require progressively less prior knowledge. These methods have also demonstrated increasing success on real-world systems. This includes emerging massively parallelized \ac{RL} in synthetic world with zero-shot transferability, \ac{IL} from diverse multimodal data sources, as well as vision-language action and world action models in robotics for acquiring generalist skills. As we scale up learning, a central question is how control-theoretic structure can more deeply inform the learning process—for example, through objective and reward design, constraint enforcement, policy parameterization, and guided exploration. Just as regularization and inductive biases are essential in supervised learning, desirable closed-loop properties—such as stability, safety, robustness, and constraint satisfaction—can serve as principled regularizers when learning is embedded in feedback. Better leveraging these properties during learning  may further improve data efficiency, reliability, and generalization.}
\item \textit{Benchmarks, open-source implementations, and reproducibility:} The field needs benchmark problems and evaluation metrics that reflect the requirements of deployment. The need for such benchmarks is becoming increasingly apparent as the discussion around the benefits of RL-oriented versus control-oriented approaches grows (e.g., \cite{song2023reaching,yuan2022safecontrolgym,weichwald2022learning}).  Such benchmarks should compare not only task performance, but also sample complexity, required prior knowledge, inference time, robustness, safety, and generalization. As discussed in~\cite{how2015benchmarks}, useful benchmarks should be sufficiently complex to expose challenges in controller design, while remaining simple enough to enable meaningful comparisons. Open-source implementations are equally important for reproducibility and cross-community progress.
\end{itemize}

\begin{figure}[t]
    \centering
    \includegraphics[width=\columnwidth]{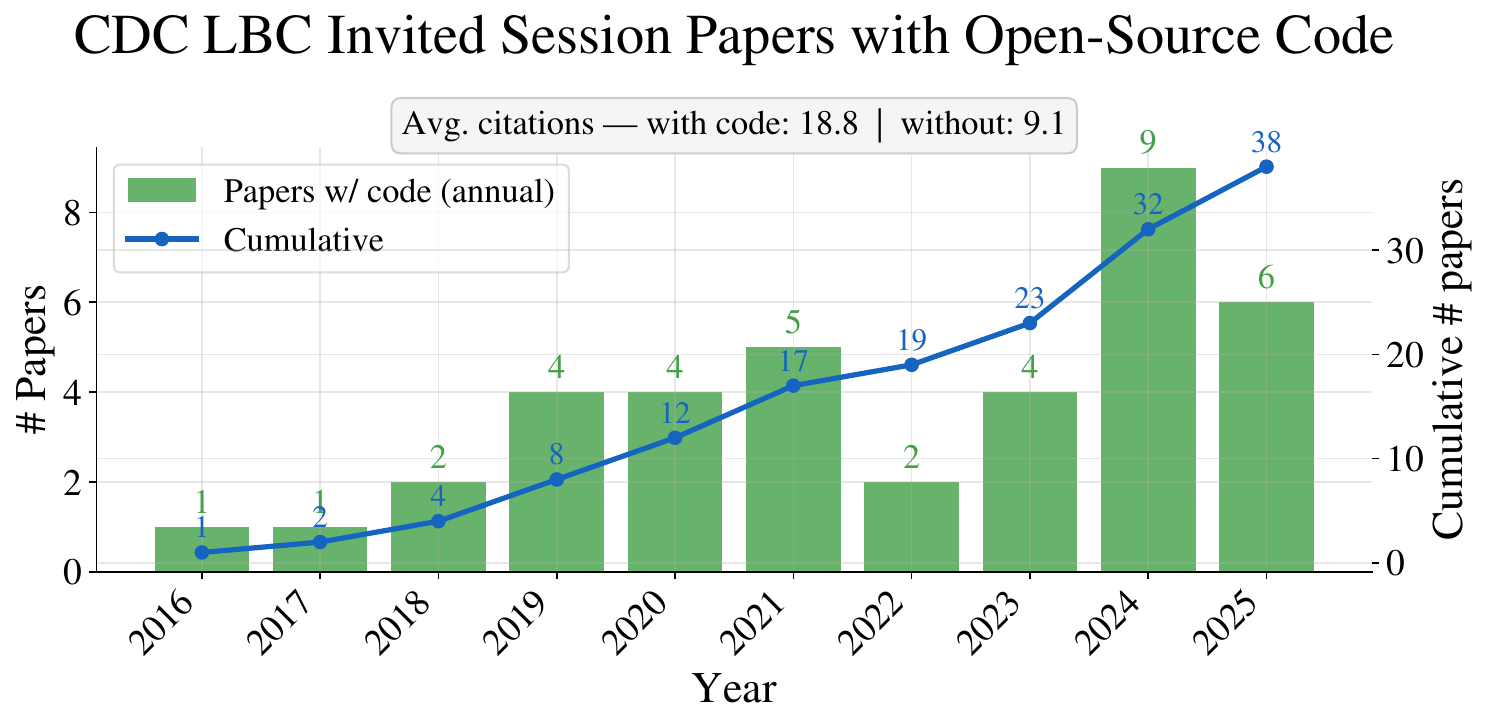}
    \caption{Number of papers that released open-source code throughout the years for the CDC Learning-Based Control Invited Sessions. In total, 38 of 182 papers, or 20.9\% of the papers, released open-source code. Details of the data source and analysis procedure are included in the appendix.}
    \label{fig:open_source_contributions}
\end{figure}

As shown in~\autoref{fig:open_source_contributions}, 20.9\% of the invited-session contributions released open-source code. This is an encouraging basis, but also indicates substantial room for improvement. As learning-based control methods become more computationally involved, communicating ideas through both theory and implementable algorithms will be essential for making new methods transparent, reproducible, and extensible.

For readers interested in contributing to the field of learning-based control, a growing collection of open-source educational materials and software tools is available, including several resources developed by the authors~\cite{ampyc,rllbc,yuan2022safecontrolgym,mooc-rl,mooc-lbc,cfr, rllbc_paper}. We hope that this tutorial provides an accessible entry point to the field and a foundation for future work toward safe, reliable, and capable autonomous systems, pushing the boundaries of learning-based control for real-world autonomy.

\section*{Acknowledgements}
APS, MA, OH, and HZ acknowledge support from the Alexander von Humboldt Foundation and the  Robotics Institute Germany (RIG)  funded by the German Federal Ministry of Education and Research (BMBF) under grant 16ME0997K.
MAM and VGL acknowledge support by the Deutsche Forschungsgemeinschaft (DFG, German Research Foundation) as part of the Research Unit ``Active Learning for Systems and Control (ALeSCo)'' – project number 535860958. 
ST and FS acknowledge support from the Robotics Institute Germany (RIG); the Deutsche Forschungsgemeinschaft (DFG, German Research Foundation) under grant TR 1433/4-1 and Germany’s Excellence Strategy EXC-2023 Internet of Production (390621612); as well as a KI-Starter grant by the state of NRW for FS. 
SZ acknowledges support from the Natural Sciences and Engineering Research Council of Canada (NSERC) through Discovery Grant RGPIN-2026-07696.

The authors thank H.~Hose, P.~Kruse, P.~Brunzema, J.~Menn, D.~Subhasish, and A.~Eisele for fruitful discussions and feedback on the manuscript, and the anonymous reviewers for helpful comments and suggestions.



\bibliographystyle{IEEEtran}
\bibliography{IEEEabrv, references, cdc_papers}

\appendix
\label{sec:appendix}
\renewcommand{\thesubsection}{\Alph{subsection}}
The dataset comprises 182 papers from the Learning-Based Control invited sessions at CDC (2016–2025). The paper list and author affiliations were compiled from the session index maintained by ETH Zurich\textsuperscript{\ref{footnote:website}}, with abstracts retrieved via the OpenAlex API and citation counts from Google Scholar. For \autoref{fig:keywords}, a fixed set of 25 domain-relevant concept phrases was determined by an LLM scan of all papers, and searched across each paper's title and abstract using case-insensitive matching. Semantically related phrases were then manually checked and consolidated into groups (e.g., reinforcement learning, deep reinforcement learning, and safe reinforcement learning are counted jointly). For \autoref{fig:open_source_contributions}, for each paper, open-source code availability was determined by scraping GitHub and project-page URLs from its abstract, then querying the GitHub API for repositories referencing the paper's arXiv ID (obtained via OpenAlex) in their README. A manual check was conducted on the existence of the repositories. Information required for \autoref{fig:learningpaperstats} for related venues (i.e., CDC, ACC, and ECC) was obtained from the DBLP API. All required information was obtained in May 2026.

\end{document}